\documentclass[10pt,letterpaper]{article}
\usepackage[top=0.85in,left=2.75in,footskip=0.75in]{geometry}

\usepackage{amsmath,amssymb}

\usepackage{changepage}

\usepackage[utf8x]{inputenc}

\usepackage{textcomp,marvosym}

\usepackage{cite}

\usepackage{nameref,hyperref}

\usepackage[right]{lineno}

\usepackage{microtype}
\DisableLigatures[f]{encoding = *, family = * }

\usepackage[table]{xcolor}

\usepackage{array}

\usepackage{graphicx}

\newcolumntype{+}{!{\vrule width 2pt}}

\newlength\savedwidth

\raggedright
\usepackage[aboveskip=1pt,labelfont=bf,labelsep=period,justification=raggedright,singlelinecheck=off]{caption}
\renewcommand{\figurename}{Fig}

\usepackage{subfig}

\makeatletter
\renewcommand{\@biblabel}[1]{\quad#1.}
\makeatother

\date{}

\usepackage{lastpage,fancyhdr,graphicx}
\usepackage{epstopdf}
\usepackage{bm}
\fancyheadoffset[L]{2.25in}
\fancyfootoffset[L]{2.25in}
\begin{document}
\vspace*{0.2in}

\begin{flushleft}
{\Large
\textbf\newline{Hierarchical communities in the walnut structure of the Japanese production network} 
}
\newline
\\
Abhijit Chakraborty\textsuperscript{1*\Yinyang},
Yuichi Kichikawa\textsuperscript{2\Yinyang},
Takashi Iino\textsuperscript{2\Yinyang},
Hiroshi Iyetomi\textsuperscript{2\Yinyang},
Hiroyasu Inoue\textsuperscript{1\Yinyang},
Yoshi Fujiwara\textsuperscript{1\Yinyang},
Hideaki Aoyama\textsuperscript{3\Yinyang}
\\
\bigskip
\textbf{1} Graduate School of Simulation Studies, The University of Hyogo, Kobe, Japan
\\
\textbf{2} Faculty of Science, Niigata University, Niigata, Japan
\\
\textbf{3} Graduate School of Science, Kyoto University, Kyoto, Japan
\\
\bigskip

%
%





* Corresponding author

Email: abhiphyiitg@gmail.com (AC)

\bigskip
\Yinyang These authors contributed equally to this work.
\end{flushleft}
\section*{Abstract}

This paper studies the structure of the Japanese production network, which includes one million firms and five million supplier-customer links. This study finds that this network forms a tightly-knit structure with a core giant strongly connected component (GSCC) surrounded by IN and OUT components constituting two half-shells of the GSCC, which we call a\textit{walnut} structure because of its shape. The hierarchical structure of the communities is studied by the Infomap method, and most of the irreducible communities are found to be at the second level. The composition of some of the major communities, including overexpressions regarding their industrial or regional nature, and the connections that exist between the communities are studied in detail. The findings obtained here cause us to question the validity and accuracy of using the conventional input-output analysis, which is expected to be useful when firms in the same sectors are highly connected to each other.




\newpage
\section*{Introduction}
A macro economy is the aggregation of the the dynamic behaviour of agents who interact with each other under diverse external (non-economic) conditions.
Economic agents are numerous and include consumers, workers, firms, financial institutions, government agencies, and countries.
The interactions of these agents result in the creation of economic networks, where nodes are economic agents, and links (edges) connect agents that interact with each other.
Therefore, there are various kinds of economic networks depending on the nature of the interactions,
which form an overlapping multi-level network of networks.
Thus, any evidence-based scientific investigation of the macro economy must be based on an understanding of the real nature of these interactions and the economic network of networks that they form.
This concept also applies to the micro-level perspective of economic agents:
without knowing who a firm trades with, how can anyone hope to determine the future of that firm?
Therefore, it is highly important to use actual network information when studying economic dynamics with either agent-based modelling/simulations or other means of systematic studies such as determining the debt-rank of an economic agent~\cite{tesfatsion2006agent,battiston2012debtrank,abergel2013econophysics,
caiani2016economics,ebook2}. 
Without this information, it is difficult to apply the validity of the results to the actual economy.

In this paper, we study the structure of one of the most important networks, the production network,
which is formed by firms (as nodes) and trade relationships (as links)
\cite{fujiwara2010large, fujiwara2012omori, Iino2015c, chakraborty2017characterization}.
In the scientific study of both the macro and the micro economy, the production network of the real economic world is a topic of high importance.
Before one engages in agent-model building and developing simulations, one needs to understand the structure of this network to be able to understand
the dynamics of this network and eventually reach into the realm of 
economic fluctuations, business cycles, systemic crises, as well as firms' growth and decline.
Therefore, in the next Section, we describe the overall statistics and visualization 
and refer to the unique overall structure of the network as a {\it``walnut''} structure.
This type of structure is quite different from what is expected because of the existence of the IN-giant strongly connected component (GSCC)-OUT components:
In the trade network, the flow of materials and goods begins with imported/mined/harvested raw materials such as oil, iron, other metals and food. Firms who engage in this business form the IN components.
These compnoents are then processed to become various products such as semiconductors or powdered food by firms, which are considered to be GSCC components,
before they are made into consumer goods by firms, which are considered to be the OUT components.
One might think that the existence of IN-GSCC-OUT components is similar to a web network 
that has a bow-tie structure \cite{broder2000gsw}. 
However, the production network is different. Ties among the firms
form a much tighter network with an overall structure that does not resemble a bow-tie.
Then, we study the community structure and reveal its hierarchical nature using the Infomap method~\cite{rosvall2008maps, rosvall2011multilevel}.

In previous studies \cite{fujiwara2010large, Iino2015c}, the modularity maximization technique~\cite{newman2004fast} is used to study the community structure of the Japanese production network. 
However, modularity maximization cannot capture the dynamic aspects of the network. This technique reveals a similar type of community partition for both directed and undirected versions of the network.
Moreover, it is well known that the modularity maximization algorithm suffers from a resolution limit problem when trying to identify the communities in a large scale network. 
The map equation method~\cite{rosvall2008maps, rosvall2011multilevel} detects communities using the dynamic behaviour of the network. In a recent study \cite{chakraborty2017characterization}, the hierarchical map equation is applied 
to characterize the level 1 communities in the Japanese production network, and a detailed investigation of the topological properties of both the intra and inter communities is conducted.  
It also shows that the regions and sectors are segregated within the communities. In another study \cite{krichene2017business}, the business cycle correlations of the communities detected by the map equation 
are studied for the network of firms listed on the Tokyo Stock Exchange. The presence of strong correlations in intra and inter communities is explained by the attributes of both the network topology and the firms.  
The crucial difference between our paper and \cite{chakraborty2017characterization, krichene2017business} is that we not only study the top level communities but also study the communities at the other levels as well as the
hierarchical structure. Moreover, we determine the compositions of the communities and subcommunities in terms of whether they include upstream and downstream firms, which has not been investigated in previous studies.  

In our paper, we conduct a level-by-level analysis and identify both communities and “irreducible” communities (communities that are not decomposed into subcommunities at the lower level).
We also study the overexpression of some of the major communities to identify both the industrial sector and the regional decomposition.
The complex nature of the links that exist between the communities are also studied.
A discussion and the conclusion as well as suggestions for future research are provided at the end. 
Some of the supporting materials are included as Appendices.
\section*{Production network data and its basic structure}\label{sec:data2} 
Our data for the production network are based on a survey conducted by Tokyo
Shoko Research (TSR), one of the leading credit research agencies in
Tokyo, and was supplied to us through the Research Institute of Economy, Trade and
Industry (RIETI). 
The data were collected by TSR by means of inquiry from firms
who represent the top five suppliers and the top five customers.
Although the large firms that have many suppliers and customers submitted replies that are incomplete, these data are supplemented with data on the other side of trade: smaller firms submit replies that include data on large firms, who are important trade partners. By combining all the submissions from both side of trade into one database, large firms are connected to
numerous smaller firms, which provides a good approximation of the real complete picture.
One might worry because some of the trades last for only a short time and sometimes they only occur once, such as when a firm seeks a good deal for just one particular occasion, and thus cast doubt on the definition of the trade network.
The form of data collection used for this study solves this problem: it is most implausible that replies containing data on a one-time trade are included, instead, data on firms that maintain a certain trade frequency are likely to be listed.
In this study, we use two datasets:
`TSR Kigyo Jouhou' (firm information), 
which contains basic financial information on more than a million
firms, 
and `TSR Kigyo Soukan Jouhou' (firm correlation information), 
which includes several million supplier-customer and ownership links
and a list of bankruptcies. 
Both of these datasets were compiled
in July 2016.
(Some of the earlier studies on the production network include~\cite{fujiwara2010large, fujiwara2012omori, Iino2015c, chakraborty2017characterization}).

In this study, $i\rightarrow j$ denotes a supplier-customer link, where firm
$i$ is a supplier for another firm $j$, or equivalently, $j$ is a
customer of $i$. We extracted only the supplier-customer links for
pairs of ``active'' firms and excluded inactive and failed firms by
using an indicator flag for them when we retrieved the basic information. We eliminated
self-loops and parallel edges (duplicate links recorded in the data),
to create a network of firms (as nodes) and supplier-customer links (as
edges). The network has the largest connected component when it is viewed as
an undirected graph, which is the giant weakly connected component
(GWCC) that includes 1,066,037 nodes (99.3\% of all the active firms)
and 4,974,802 edges. 

This study not only analyzes the network but considers several attributes of each node: the financial information in terms of firm size, which is measured as
sales, profit, number of employees and the firm's growth; the major and minor
classifications of industrial sectors, details regarding the firm's products, the
firm's main banks, the principal shareholders, and miscellaneous other
information including geographical location. For the purpose of our study,
we focus on two attributes of each firm, namely the industrial sector
and the geographical location of the head office.

\begin{figure}[ht]
    \begin{center}
   \includegraphics[width=\textwidth]{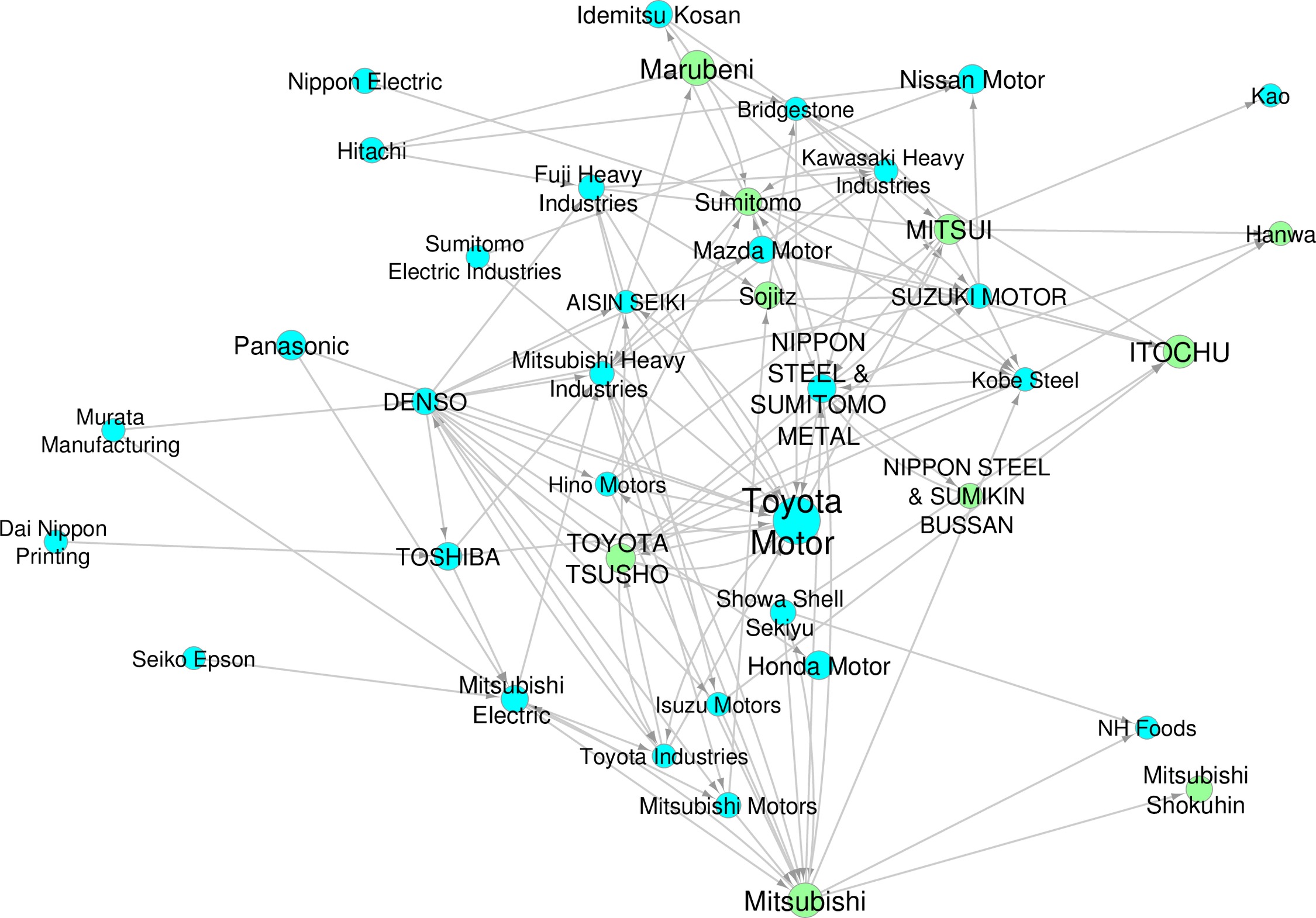} 
    \end{center}
    \caption{
    {\bf Representative network of the automobile industry in Japan.} Major firms are selected under the following conditions: i) they are connected to Toyota
    Motor within three degrees of separation, ii) they belong to either the manufacturing or wholesale sectors, iii) they are listed in the first section of the Tokyo Stock Exchange, and
    iv) They are in the top 40 in terms of sales. The firms thus selected are displayed as nodes and the transactions between them are displayed as arrows. All of the displayed nodes 
    belong to the GSCC component. The size of the nodes is scaled to the sales of the corresponding firm. The color of the nodes distinguishes
    their industry type; blue and green designate manufacturing and wholesale, respectively.
    }
    \label{fig:majorfirms}
    \end{figure}

The industrial sectors are hierarchically categorized into 20 divisions,
99 major groups, 529 minor groups and 1,455 
industries
(Japan Standard Industrial Classification, November 2007, Revision~12).
See 
Table A 
in S1 Appendix
for the number of firms in each division of each industrial sector. 
Each firm is 
classified according to the sector it belongs to, and the primary,
secondary and tertiary, if any, is identified.
The geographical location is converted into a level of one of
47 prefectures or into one of 9 regions
(Hokkaido, Tohoku, Kanto, Tokyo, Chubu, Kansai, Chugoku, Shikoku, and Kyushu). 
See 
Table B 
in S1 Appendix
for the number of firms in each regional area of Japan.
Fig~\ref{fig:majorfirms} depicts a representative supply-chain network of the automobile industry in Japan. 
For example, Toyota Motor Corporation, the largest car manufacturer in the nation, obtains mechanical parts from 
suppliers such as Denso and Aisin Seiki. In addition, Toyota is indirectly connected to Denso through Aisin Seiki. 
One can also go up from Denso to Murata Manufacturing in the figure. For electronic parts, another 
important components of cars, Toyota has direct transactions with general electrical manufacturers such as
Toshiba and Panasonic, and Toshiba, in turn, obtains parts from Dai Nippon Printing. General trading companies
such as Marubeni, Mitsui, and Toyota Tsusho play a key role in the formation of the supply-chain network. 
In addition, we can observe a circular transaction relation among Toyota Motor, Denso, and Toyota Industries. 
The existence of such a feedback loop can complicate firms' dynamics in the production network.

In terms of the flow of goods and services (and money in the reverse direction),
the firms are classified in three categories: the ``IN" component, the ``GSCC", and the ``OUT" component.
This structure is called ``bow-tie'' in a well-known study on the Internet \cite{broder2000gsw}.
The GWCC can be decomposed into the parts defined as follows:
\begin{description}%
\setlength{\itemsep}{0pt}
\item[GWCC] the giant weakly connected component: the largest connected
  component when the network is viewed as an undirected graph. An undirected path
  exists for each arbitrary pair of firms in the component.
\item[GSCC] the giant strongly connected component: the largest connected
  component when the network is viewed as a directed graph. A directed path
  exists for each arbitrary pair of firms in the component.
\item[IN] The firms through which the GSCC is reached via a direct
  path.
\item[OUT] The firms that are reachable from the GSCC via a direct
  path.
\item[TE] ``Tendrils''; the remainder of the GWCC
\end{description}
It follows from the definitions that
\begin{equation}
  \label{eq:walnut}
  \text{GWCC}=\text{GSCC}+\text{IN}+\text{OUT}+\text{TE}
\end{equation}


We, however, find it far more appropriate to call this structure a `'Walnut" structure,
as ``IN" and ``OUT" components are not as separated as in the
two wings of a ``bow-tie" but are more like the two halves of a walnut shell,
surrounding the central GSCC core. 
This can be explained as follows.
The number of firms in each component of the GSCC, IN, OUT and TE is shown
in Table~\ref{tab:walnut_num}. 
Half of the firms are inside the GSCC.
20\% of the firms are in the upstream side or IN, and
26\% of them are in the downstream side or OUT.

\begin{table}[ht]
\centering
\caption{\bf Walnut structure: The sizes of the different components}
\begin{tabular}{c|r|r}
  \hline
  Component & \#firms & Ratio (\%) \\
  \hline 
  GSCC & 530,174 & 49.7  \\
  IN & 219,927 & 20.6  \\
  OUT & 278,880 & 26.2  \\
  TE & 37,056 & 3.5  \\
  \hline
  Total & 1,066,037 & 100 \\
  \hline
\end{tabular}
\begin{flushleft}
``Ratio" refers to the ratio of the number of firms to the total number of the firms in the GWCC.
\end{flushleft}
\label{tab:walnut_num}
\end{table}

\begin{figure}[ht]
\centering
    \includegraphics[width=0.95\textwidth]{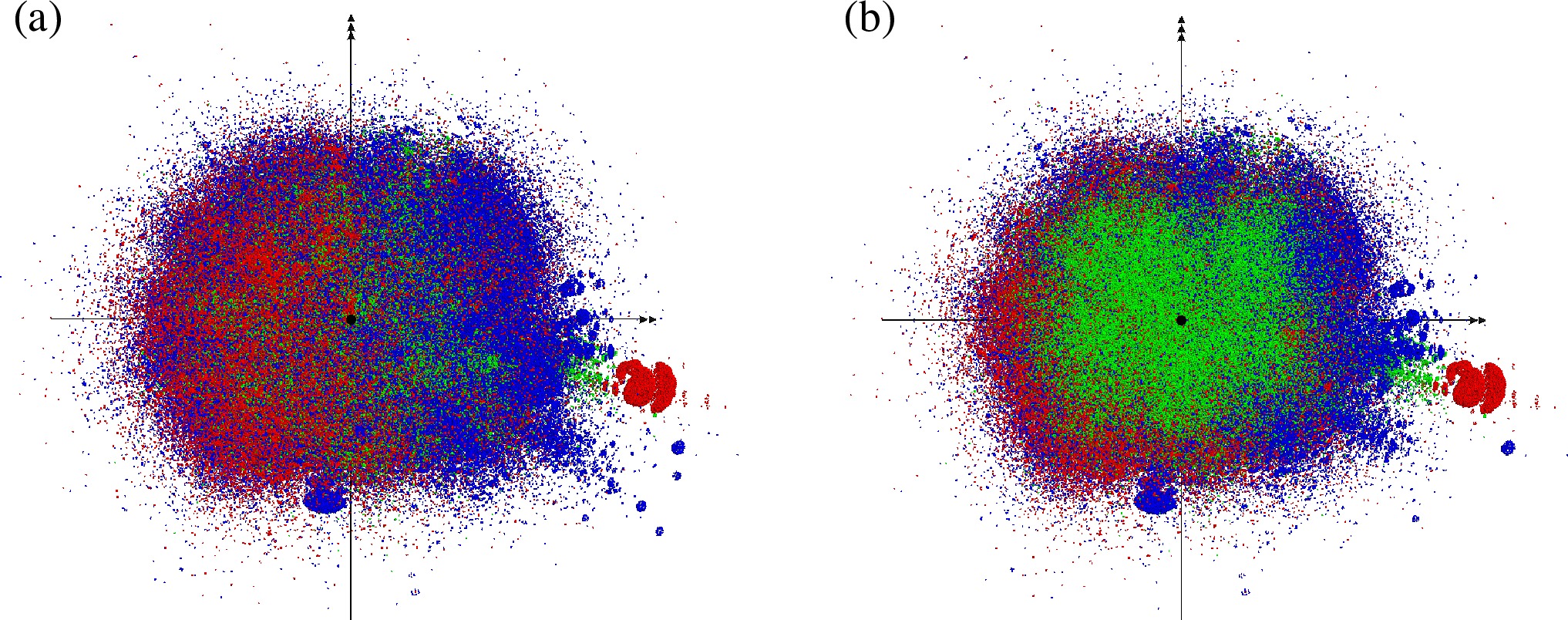}
    \caption{{\bf Visualization of the network in three-dimensional space}
         A surface view of the network is shown in panel (a), and a cross-sectional view that is cut through its center is shown in panel (b). The red, green, and 
         blue
         dots represent firms in the IN, GSCC, and OUT components, respectively.}
    \label{fig:walnut}
\end{figure}

In contrast with the well-known ``bow-tie structure'' in the study conducted by \cite{broder2000gsw}
(in which the GSCC is less than one-third of the GWCC),
the GSCC in the production network occupies half of the system,
meaning that most firms are interconnected by the small geodesic distances
or the shortest-path lengths in the economy.
In fact, by using a standard graph layout algorithm based on
a spring-electrostatic model with three-dimensional space \cite{SPE:SPE4380211102},
we can show in Fig~\ref{fig:walnut} by visual inspection
how closely most firms are interconnected with each other.

\begin{figure}[ht]
\begin{center}
    \includegraphics[width=0.4\textwidth]{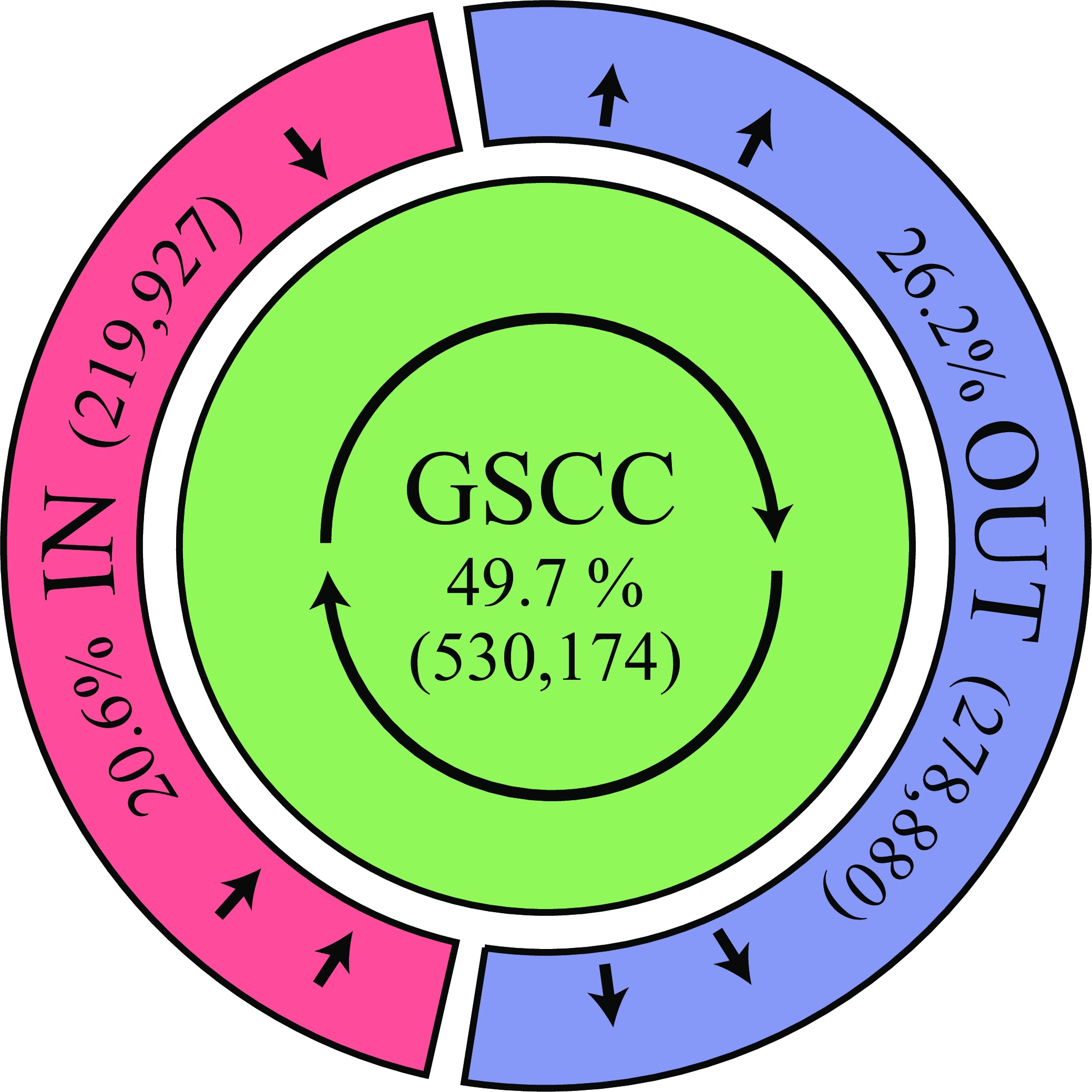}
\end{center}
    \caption{{\bf The Walnut structure.}
  The production network as a walnut structure. 
  The area of each component is approximately proportional to its size.}
    \label{fig:walnut2}
\end{figure}

Moreover, by examining the shortest-path lengths from GSCC to
IN and OUT as shown in Table~\ref{tab:walnut_dist}, one can
observe that the firms in the upstream or downstream sides
are mostly located a single step away from the GSCC.
This feature of the economic network is different from
the bow-tie structure of many other complex networks.
For example, the hyperlinks between web pages of a similar size,
(GWCC: 855,802, GSCC: 434,818 (51\%), IN: 180,902 (21\%), OUT: 165,675 (19\%), TE: 74,407 (9\%))
which are studied in \cite{leskovec2009cs}, have a bow-tie structure such that
the maximum distance from the GSCC to either IN or OUT is 17,
while more than 10\% of the web pages in IN or OUT are located
more than a single step away from the GSCC.
This observation as well as Fig~\ref{fig:walnut}
leads us to say that the production network has a ``walnut'' structure,
rather than a bow-tie structure. We depict the schematic diagram
in Fig~\ref{fig:walnut2}

Later, we shall show how each densely connected module or community
is located in the walnut structure. 

\begin{table}[ht]
\centering
\caption{\bf Walnut structure: the shortest distance from GSCC to IN/OUT}
\begin{tabular}[c]{crr|crr}
  \hline
  \multicolumn{3}{c|}{IN to GSCC} &
  \multicolumn{3}{c}{OUT to GSCC} \\ \hline
  Distance & \#firms&Ratio (\%)& Distance & \#firms &Ratio (\%)\\
  \hline
  1 & 212,958 &96.831&
  1 & 266,925 &95.713\\
  2 & 6,793 & 3.089&
  2 & 11,650 &4.177\\
  3 & 170 &0.077&
  3 & 296 &0.106\\
  4 & 6 &0.003&
  4 & 9 &0.003\\
  \hline
  Total & 219,927 &100&
  Total & 278,880 &100\\
  \hline
\end{tabular}
\begin{flushleft}
The left half shows the number of firms in the IN component that connects
to the GSCC firms with the shortest distance 1--4. The left side shows the OUT component. 
\end{flushleft}
\label{tab:walnut_dist}
\end{table}
\clearpage
\section*{Methods}
\subsection*{Community detection}
Community detection is widely used to elucidate the structural properties of large-scale networks.
In general, real networks are highly non-uniform.
Community detection singles out groups of nodes densely connected to each other in a network to divide that network into modules.
This process enables us to have a coarse-grained view of the structure of such complicated networks.
One of the most popular methods used for community division is maximizing the modularity index\cite{newman2004fast}.
Modularity measures the strength of the partition of a network into communities by comparing the fraction of links in given communities with the expected fraction of links if links were randomized with the same degree of distribution as the original network. 
However, it is well known that the modularity method suffers from a problem called resolution limit\cite{fortunato2007resolution} when applied to large networks. That is, optimizing modularity fails to detect small communities even if they are well defined, such as cliques.

The map equation method~\cite{rosvall2008maps} is another method used to detect communities in a network. 
This method is found to be one of the best performing community detection techniques compared to the others~\cite{lancichinetti2009community}. 
The map equation method is a flow-based and information-theoretic method depending on the map equation, which is
defined as
\begin{equation}
L(C)=q_\curvearrowright H(C)+\sum^m_{i=1}p^i_\circlearrowright H(\mathcal{P}^{i})\ .
\label{mapeq}
\end{equation}
Here, $L(C)$ measures the per step average description length of the dynamics of a random walker migrating through the links between the nodes of a network with a given node partition $C=\{C_{1},\cdots,C_{\ell}\}$ that consists of two parts. The first term arises from the movements of the random walker across communities, where $q_\curvearrowright$ is the probability that the random walker switches communities, and $H(C)$ is the average description length of the community index codewords given by the Shannon entropy. The second term arises from the movements of the random walker within the communities, where $p^i_\circlearrowright$ is the percentage of the movements within the community $C_{i}$, and $H(\mathcal{P}^i)$ is the entropy of the codewords in the module codebook $i$.

If the network has densely connected parts in which a random walker stays a long time, one can compress the description length of the random walk dynamics in a network by using a two-level codebook 
for nodes adapted to such a community structure; this is similar to geographical maps in which different cities recycle the same street names such as ``main street'~\cite{rosvall2008maps}. Therefore, obtaining the best community decomposition in the map equation framework amounts to searching for the node partition that minimizes the average description length $L(C)$.

In regard to the resolution limit problem, any two-level community detection algorithms including the map equation are not able to eliminate the limitation. However, the map equation significantly mitigates the problem as has been shown by a recent theoretical analysis~\cite{PhysRevE.91.012809}. In practice, this is true for our network, as will be demonstrated later. 

Recently, the original map equation method has been extended to networks with multi-scale inhomogeneity.
A network is decomposed into modules that include their submodules and then their subsubmodules and so forth.
The hierarchical map equation\cite{rosvall2011multilevel} recursively searches for such a multilevel solution by minimizing the description length with possible hierarchical partitions. The map equation framework for the community detection of networks is now more powerful. 
Therefore, we analyze the production network using this method. 
The code of the hierarchical map equation algorithm is available at http://www.mapequation.org.

Note that this study exclusively considers the community identification for nodes in our network. That is, each node belongs to a unique community at every hierarchical level. However, such community assignment may be too restrictive for a small number of giant conglomerate firms such as Hitachi and Toshiba because of the diversity of their businesses. The map equation is so flexible that it can detect the overlapping community structure of a network in which any node can be a member of multiple communities~\cite{PhysRevX.1.021025}. However, we use the original algorithm as an initial step toward obtaining a full account of the firm-to-firm transaction data.

\subsection*{Overexpression within communities and subcommunities}
\label{Method:overexpression}
Most real-world networks have a community structure~\cite{fortunato2010community}.  Such communities are formed in a network based on the principle of homophily~\cite{currarini2009economic}. This principle indicates that a node has a tendency to connect with other similar nodes. For example, ethnic and racial segregation are observed in our society~\cite{echenique2007measure}, biological functions play a key role in the formation of communities in 
protein-protein interaction networks~\cite{chen2006detecting}, and the community structure of stock markets is similar to that of their economic sectors~\cite{onnela2003dynamics}.
We find that attributes play a crucial role in the formation of the community structure of the production network using the following method.  

We follow the procedure used in~\cite{tumminello2011community} to determine the statistically significant overexpression of different locations and sectors within a community. This method was developed from the statistical validation of the overexpression of genes in specific terms of the Gene Ontology database~\cite{druaghici2003data}.
In this procedure, a hypergeometric distribution $H(X|N, N_C, N_Q)$ is used to measure the probability that $X$ randomly selected nodes in community $C$ of size $N_C$ will have attribute $Q$. The hypergeometric distribution $H(X|N, N_C, N_Q)$ can be written as
\begin{equation}
H(X|N, N_C, N_Q)= \frac{\binom{N_C}{X}\binom{N-N_C}{N_Q-X}}{\binom{N}{N_Q}}\ ,
\end{equation}
where $N_Q$ is the total number of elements in the system with attribute $Q$.
Further, one can associate a {\em p value} $p(N_{C,Q})$ with $N_{C,Q}$ nodes, having attribute $Q$ in community $C$ with $H(X|N, N_C, N_Q)$ by the following relation:
\begin{equation}
p(N_{C,Q}) = 1- \sum_{X=0}^{N_{C,Q}-1}H(X|N, N_C, N_Q)\ .
\end{equation}
The attribute $Q$ is overexpressed within community $C$ if $p(N_{C,Q})$ is found to be lower than some threshold value $p_c$.
As we use a multiple-hypothesis test, we need to choose $p_c$ appropriately to exclude false positives.
We assume that $p_c=0.01/N_A$, as specified in~\cite{tumminello2011community}, which includes a Bonferroni correction~\cite{miller1981normal}. Here, $N_A$ represents the total number of different attributes 
(In our study we have $N_A=9$ regional attributes) for all the nodes of the system.
\clearpage
\section*{Results}
\subsection*{Hierarchy of communities}

\begin{figure}[hbt]
\begin{center}
\includegraphics[width=0.8\textwidth]{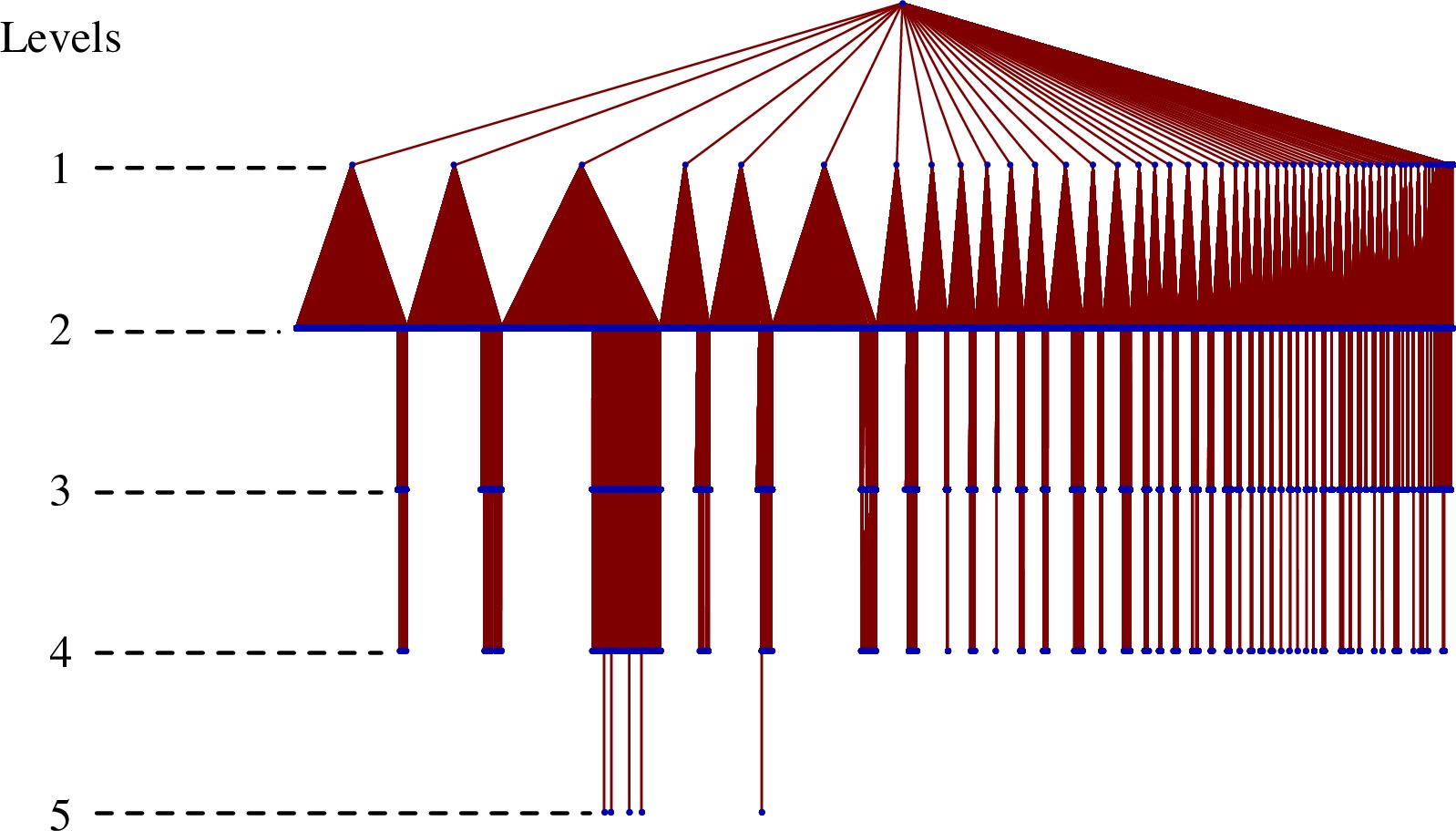}
\end{center}
\caption{{\bf Hierarchical structure of the communities}
Five levels of hierarchical community decomposition are illustrated. 
The width of the triangle originating in each community at the $n$-th level is proportional to the number of its subcomunities at the $(n+1)$-th level.}
\label{fig:hia}
\end{figure}

\begin{figure}[!h]
\begin{center}
\includegraphics[width=\textwidth]{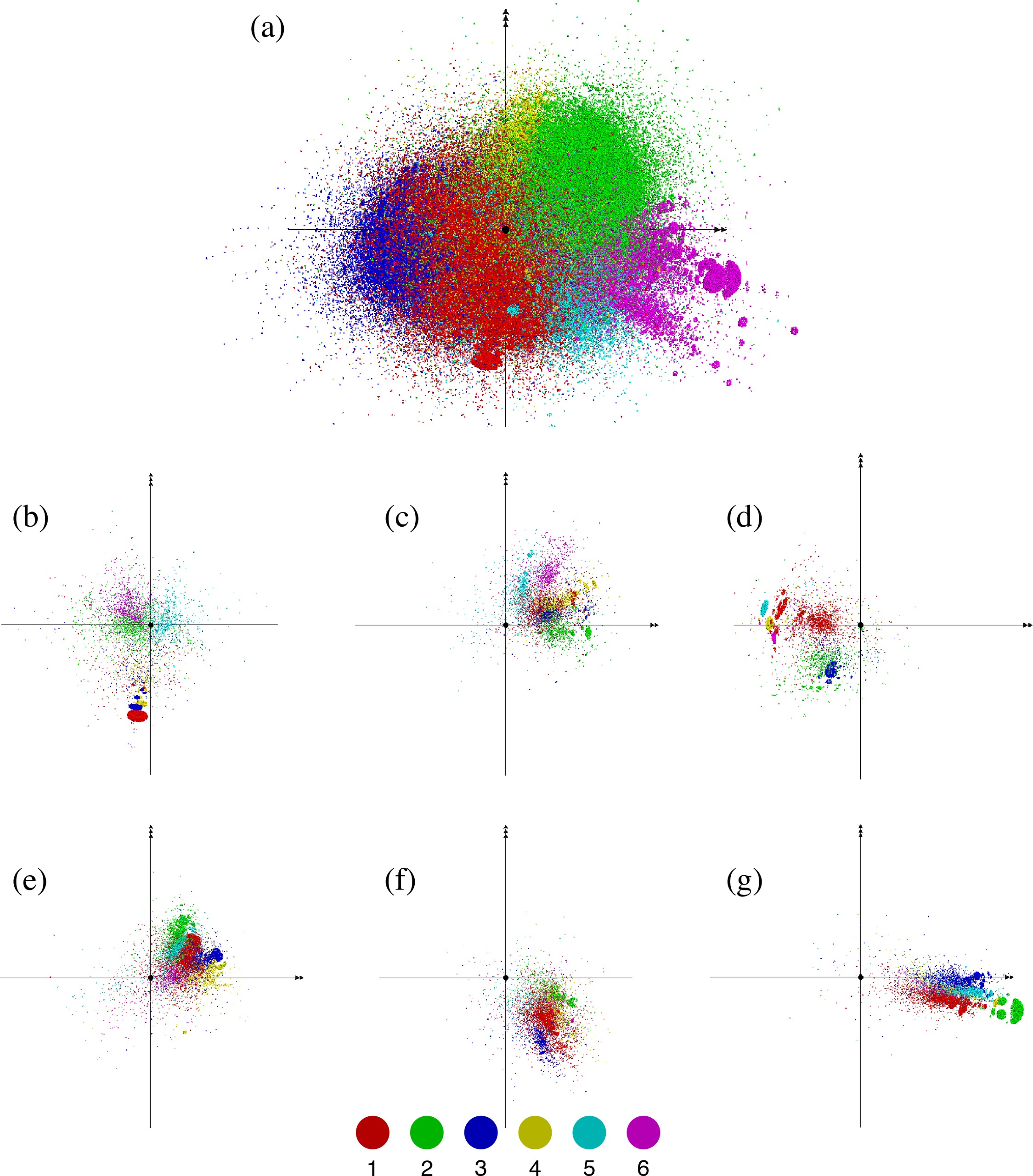}
\end{center}
\caption{{\bf Hierarchical decomposition of the whole network into communities and subcommunities.}
This panel (a) highlights the 
6
largest communities at the top modular level with different colors. Each of these communities is further decomposed into subcommunities as demonstrated in panels (b) through (g), where the 6th largest subcommunities of the 1st through the 6th largest communities are highlighted.}
\label{fig:comp_com_decomp}
\end{figure}

By using the Infomap method \cite{rosvall2008maps, rosvall2011multilevel}, 
we find that the communities have a hierarchical structure, as summarized in Table \ref{tab:hia}, and determine the number of firms at each level. 
This hierarchical structure is illustrated in Fig~\ref{fig:hia}, where
2nd level communities are lined up from left to right in
a descending order in terms of community size (number of firms),
and the width of the triangles reflects the number of subcommunities in each
community.
We find that most of the subcommunites are on the 2nd level
and that most of the firms (94\%) belong to 2nd level communities.
Compared with 1st and 2nd level communities, the 3rd to the 5th levels are of no 
significant importance. Therefore, we
limit our discussion of the properties of the (sub)communities to those of the 
2nd level.
Past studies on the application of the hierarchical map equation to real world networks~\cite{rosvall2011multilevel, PhysRevE.91.012809} show that 
dense networks have large communities at the finest level with shallow hierarchies, and 
sparse networks tend to have deep hierarchies. It is also observed that the depth 
of the hierarchies increases with network size. In the case of the California road network, the hierarchy has a deep level because the road network has geographical constraints that decrease the number of shortcuts between the different parts of the network~\cite{rosvall2011multilevel}.
In our production network, we observe a relatively shallow hierarchy because it does not
 have such strict constraints.      

\begin{table}[hbt]
\centering
\caption{\bf Modular level statistics}
\begin{tabular}{c|r|r|r|r}
\hline
Level & \#com & \#irr.com  & \#firms& Ratio (\%)\\ \hline
$1$  &$209$ & $106$  & 830 &0.078\\ \hline
$2$  &$65,303$ &$60,603$ & 998,267 & 93.643 \\ \hline
$3$  &$18,271$ &$17,834$ & 61,748 & 5.792\\ \hline
$4$  &$1,544$ & 1,539  & 5,168 &0.485 \\ \hline
$5$  &$10$ &$10$  &  24 & 0.002\\ \hline
Total &&80,092&1,066,037 &100.00\\ \hline
\end{tabular}
\begin{flushleft}
Results of community detection using the multi-coding Infomap method.
``\#com" is the number of all the communities,
``\#irr.com'' is the number of irreducible communities, which are communities
that do not have any subcommunities.
``\#firms" refers to the number of firms in irreducible communities
\end{flushleft}
\label{tab:hia}
\end{table}

 We visualize the hierarchical decomposition of the whole network into communities and their subcommunities in Fig~\ref{fig:comp_com_decomp}. The configuration of the nodes in three-dimensional space is the same as that in Fig~\ref{fig:walnut}. We can see that the network is extremely complex with multi-scale inhomogeneity. The results of an overexpression analysis indicate that the major communities of the 1st and 2nd levels are characterized as industrial sectors and regions, as noted in the subsequent subsections.

For the purpose of making the following discussion of communities
transparent, let us adopt the following indexing convention:
At the top modular level of the hierarchical tree structure,
the communities are indexed by their rank in size (the number of firms
in the community). Thus, the largest
community at the top level is denoted as ``C$_1$''.
At the lower levels, the rank of the size is added after `:'. 
For example, community ``C$_{1:5}$'' is 
the fifth largest 2nd level community among all the 2nd-level communities 
that belong to the largest top-level community C$_1$.



\subsection*{Level-1 communities}
\begin{figure}[!h]
\begin{center}
\includegraphics[width=0.7\textwidth]{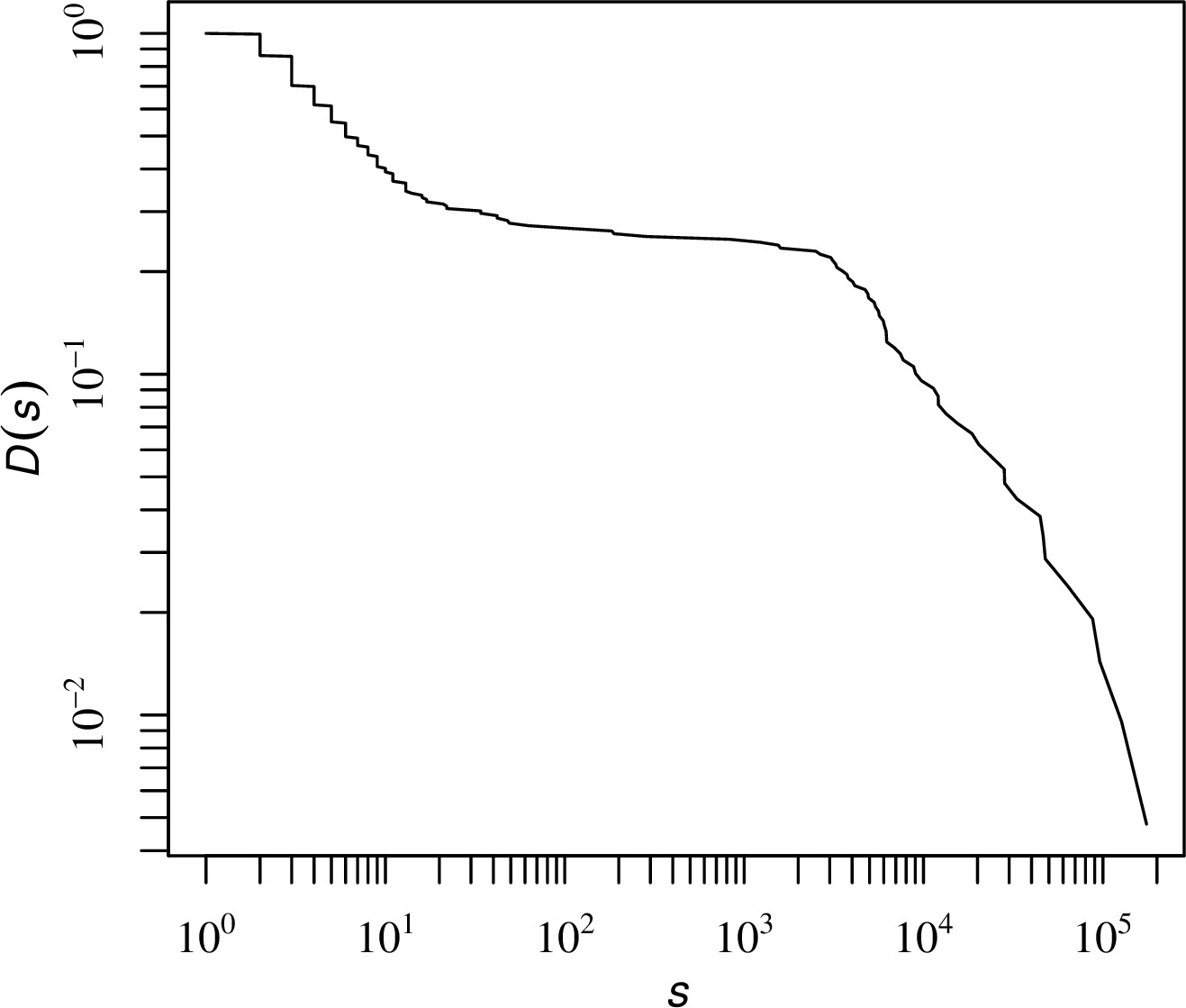}
\end{center}
\caption{
{\bf The complementary cumulative distribution function $D(s)$ of the community size $s$ at the top modular level.}
}
\label{fig6}
\end{figure}

 The complementary cumulative function {\em D(s)} indicates the fraction of communities at the top level having a size of at least s, as shown in Fig~\ref{fig6}.
The bimodal nature of the distributions manifests the resolution limit problem.
A small number of communities predominates the whole system. Among some 200 communities detected, for example, the largest communities contain 100,000-200,000 firms. However, such extremely large communities are decomposed into subcommunities by the hierarchical map equation in a unified way. This process is quite different from community detection based on modularity. One may address this problem by applying the modularity maximization method recursively; communities are regarded as separated subnetworks that can be further decomposed. However, this procedure lacks a sound basis because it uses different null models to decompose the subnetworks~\cite{fortunato2010community}.
A more detailed comparison between these two methods is provided in S1 Appendix.

\begin{figure}[!h]
    \begin{center}
    \includegraphics[width=0.7\textwidth]{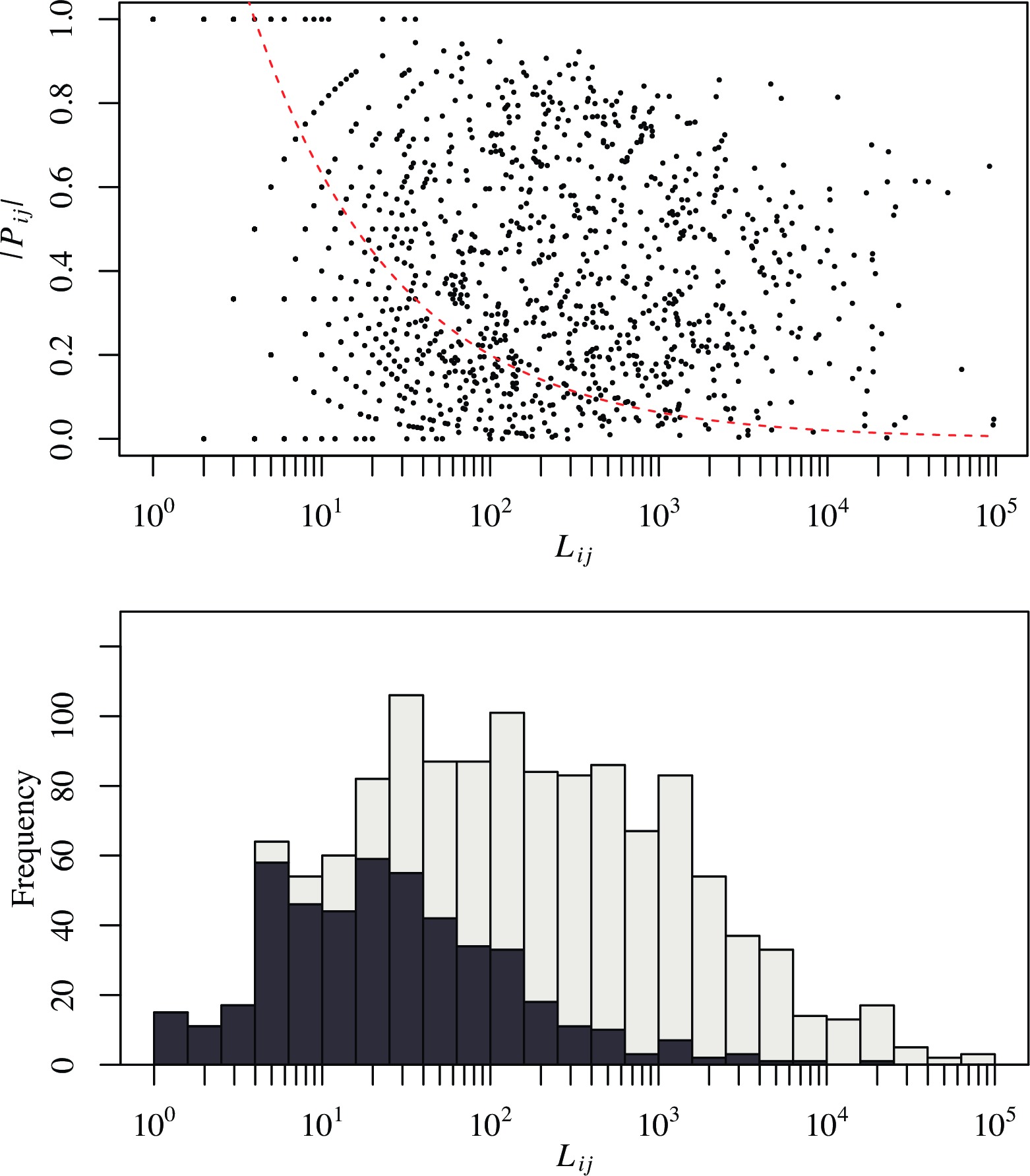} 
    \end{center}
    \caption{
    {\bf Polarizability of the direction of links interconnecting communities at the top level.} Here, 51 major communities containing more than 1,000 firms are selected.
    The top figure plots the polarization ratio $|P_{ij}|$ of the linkage between communities $i$ and $j$ versus the total number $L_{ij}$ of its constituting links.
    The dashed curve shows the significance level corresponding to $2\sigma$ for the polarizability of intercommunity linkage for the given total number of its constituents, where the random orientation of the individual links is adopted as a null model; see Eq. (\ref{eq:sigma}) for the standard deviation $\sigma$.
    The bottom figure is a histogram for the frequency of intercommunity linkages in each bin of $L_{ij}$.
    The grey (black) bars depict the number of intercommunity linkages with a $|P_{ij}|$ that is higher (lower) than the threshold for the test of statistical significance.}
    \label{fig:poltop}
\end{figure}

The map equation is a method that can be used to divide a directed network into communities in which nodes are tightly connected in both directions. Due to the nature of the network, the flows across communities thus detected should be biased in an either direction.
Fig~\ref{fig:poltop} confirms this expectation. To quantify the polarizability of the links between a pair of communities,
we introduce the polarization ratio defined by 
\begin{align}\label{eq:polarization}
P_{ij} = \frac{A_{ij}-A_{ji}}{A_{ij}+A_{ji}},
\end{align}
where $A_{ij}$ is the total number of links spanning from communities $i$ to $j$ and $A_{ji}$ and that of the opposite links.
If the linkage between communities $i$ and $j$ is completely polarized, then $P_{ij}$ becomes $\pm 1$ depending on its direction; if the linkage is evenly balanced, then $P_{ij}=0$. If we assume that the links have no preference with respect to their direction as a null hypothesis, then the null model predicts that the polarization ratio for the connections between communities $i$ and $j$ fluctuates around 0 with the standard deviation $\sigma$ given by
\begin{equation}
\sigma=\frac{1}{\sqrt{L_{ij}}}\ ,
\label{eq:sigma}
\end{equation}
where $L_{ij} = A_{ij}+A_{ji}$ is the total number of links between the two communities. 
If we focus on intercommunity linkages with $L_{ij}\geq 100$, we see that the ones whose direction is polarized in a statistically meaningful way occupy $86.7\%$ of their total. The corresponding share of intercommunity linkages is $70.1\%$ for $L_{ij}\geq 10$.
Most of the connections between communities with more than 100 links are significantly polarized in reference to the random orientation model for intercommunity links.


We find the overexpression of the attributes in 1st level communities to determine the factors that play a crucial role in the formation of such communities. Our study considers both the location and the sector attributes. The location attributes are divided into $9$ regions, and the sector attributes are categorized in $20$ divisions. 
The details about the sixth largest 1st level communities and the overexpressed attributes within it are tabulated in Table~\ref{table_1st_modular}. We also use a finer classification, i.e., $47$ prefectures and $99$ major sectors for which the results are provided in S1 Appendix.  
We observe a strong connection between overexpressed sectors and overexpressed regions. In the largest community, mainly manufacturing sectors and heavily urbanized regions (Kanto,
Tokyo, Chubu, and Kansai) are overexpressed. The 2nd largest community shows that mainly the agriculture and food industries (see SI) and rural regions (Hokkaido, Tohoku,
Shikoku, and Kyusyu-Okinawa) are overexpressed. In terms of overexpression in the 3rd largest community, the construction sector dominates and the corresponding overexpressed region indicates these firms are mainly
based in Kanto and Tokyo. The transport and wholesale retail trade industries are the dominate attributes of the 4th largest community, and Tohoku, Kanto, and Chubu are the overexpressed regions.
The 5th largest community mainly includes Tokyo, and the primary overexpressed sectors are information and communications, scientific research, and professional and technical services. The 6th largest community primarily primarily includes medicine and health care. To summarize, the following characterizes the six largest communities:

\begin{itemize}
    \item The largest community: Manufacturing sectors
    \item The second largest community: Food sectors
    \item The third largest community:  Construction sectors 
    \item The fourth largest community: Wholesale and retail trade 
    \item The fifth largest community:  IT sector and scientific research, primarily based in Tokyo 
    \item The sixth largest community: Medical and health care
\end{itemize}


\begin{figure}[!h]
\begin{center}
\includegraphics[width=\textwidth]{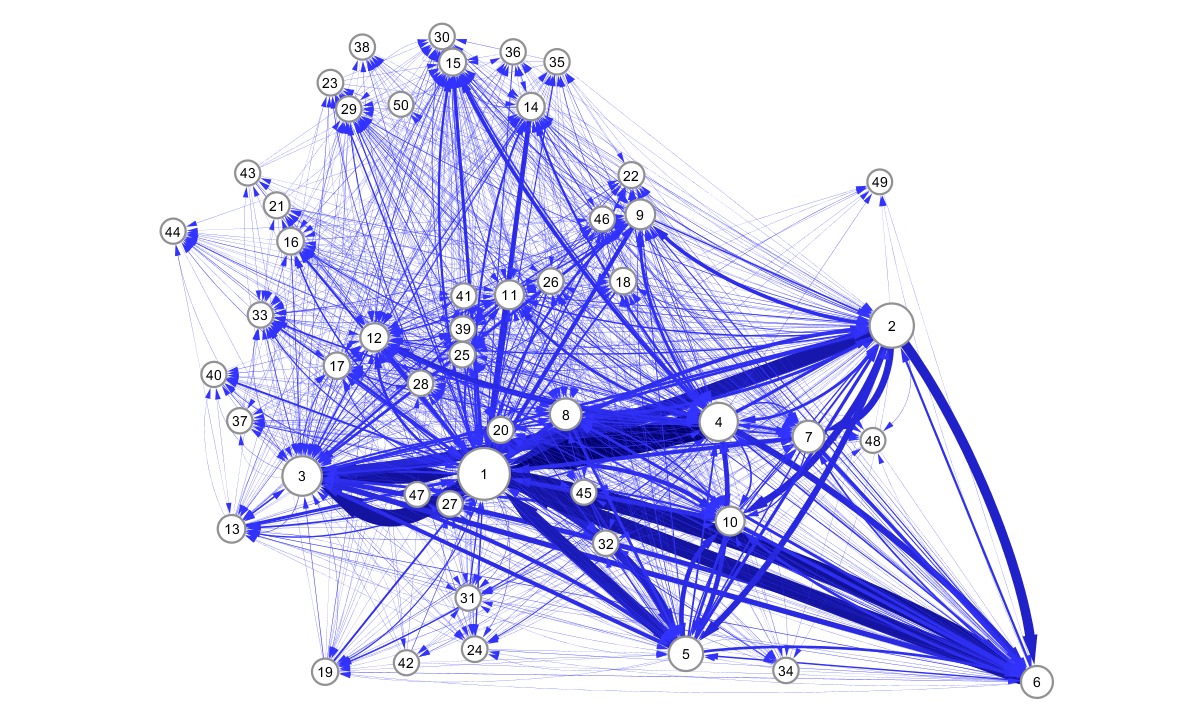}
\end{center}
\caption{{\bf Network of the 50 largest communities at the top level.} The major communities are depicted as nodes, and their size is scaled to the size of their corresponding communities. A bundle of directed links connecting a pair of nodes in either direction is represented by an arrow, the width of which is proportional to the total number of their links.}\label{fig:flowdiagram}
\end{figure}

\begin{table}[!h]
\small 
\centering
\caption{\bf Overexpressions of the 1st level communities}
\scalebox{0.75}{
\begin{tabular}{|c|r|r|p{4.0cm}|p{5.0cm}|ccc|}
  \hline
Index & Size & \#subcom & Region & Sector & IN & GSCC & OUT \\ 
  \hline
1 & 175,150 & 7135 & Kanto (0.21);\newline Tokyo (0.14);\newline Chubu (0.22);\newline Kansai (0.21) 
& Manufacturing (0.33); & 0.20 & 0.65 & 0.14 
\\ \hline
2 & 126,997 & 5455 & Hokkaido (0.07);\newline Tohoku (0.11);\newline Shikoku (0.05);\newline Kyusyu-Okinawa (0.13)
& Agriculture (0.04); 
Manufacturing (0.18); Wholesale and retail (0.43); Accommodations (0.11); Living-related (0.03); Compound services (0.02) & 0.11 & 0.46 & 0.40
\\ \hline
3 & 96,062 & 7339& Kanto (0.48);\newline Tokyo (0.25) & 
Construction (0.64);  
Real estate (0.09);  Scientific research (0.06); & 0.39 & 0.38 & 0.16
\\ \hline
4 & 87,647 & 2660& Tohoku (0.11);\newline Kanto (0.22);\newline Chubu (0.20) 
& Transport (0.15); Retail (0.38); Finance (0.05); 
Services, N.E.C. (0.17) & 0.11 & 0.43 & 0.44
\\ \hline
5 & 63,611 & 3631& Tokyo (0.40)
& Information (0.25); Finance (0.01); Real estate (0.05); Scientific research (0.13); Living-related (0.05);  Education (0.01); Services, N.E.C. (0.07) & 0.26 & 0.45 & 0.26\\ 
\hline
6 & 47, 759 & 6214& Hokkaido (0.06);\newline Tokyo (0.22); \newline Chugoku (0.08); \newline Shikoku (0.05); \newline Kyusyu-Okinawa (0.13)  
& Wholesale and retail (0.28); Living-related (0.05); Medical (0.48) & 0.24 & 0.21 & 0.52\\ 
\hline
\end{tabular}
}
\begin{flushleft} 
``\#subcom'' is the total number of subcommunities included in each of the 1st level communities. The overexpression in terms of the regions and sector-divisions of the 6th largest communities at the 1st level.
The percentage of nodes having a particular attribute is indicated in parentheses.
Those with less than 0.01 are not listed. In addition, the percentages of the IN, GSCC, and OUT components are listed for each community.
\end{flushleft}
\label{table_1st_modular}
\end{table}

Fig~\ref{fig:flowdiagram} is a coarse-grained diagram of the network shown in Fig.~\ref{fig:walnut}, where the 50 largest communities at the top level are represented by nodes, and the direct links connecting them, in either direction, are bundled into arrows. We used the following steps to prepare the diagram. We first calculated the center of mass for the IN, GSCC, and OUT components in three-dimensional space. The three centers thus obtained determine the two-dimensional plane for the drawing. Second, we fixed the horizontal axis to optimally represent the direction of flow from the IN (left-hand side) components to the OUT (right-hand side) components through the GSCC; in fact, the three centers are almost aligned horizontally. Then, we calculated the center of mass of the major communities and projected them onto the two-dimensional plane to layout the major communities onto it. Finally, we connected these communities by arrows using information on the links between them.

The positions of the communities on the horizontal line clearly reflect their characteristics in terms of the walnut structure, as shown in Table~\ref{table_1st_modular}. Among the 6 largest communities, the 3rd community contains twice as many IN components as the averaged concentration on the leftmost side. On the other hand, the 6th community with the largest OUT concentration is on the rightmost side. The 2nd and 4th communities, which are dominated by OUT components, are also on the right-hand side. The 1st community with excess GSCC components is between the 3rd community and the OUT-excess communities. The 5th community, whose composition is very close to the average one, is rather in middle of the walnut structure. Most of the remaining relatively small communities are localized on the left-hand side. This configuration is understandable, because the IN and GSCC components tend to form integrated communities, as will be shown later.


\subsection*{Level-2 communities}
At the 2nd level, some of the top level communities are decomposed to 
several subcommunities as shown 
in Tables D and E
in S1 Appendix.

\begin{figure}[!h]
\begin{center}
\includegraphics[width=0.7\textwidth]{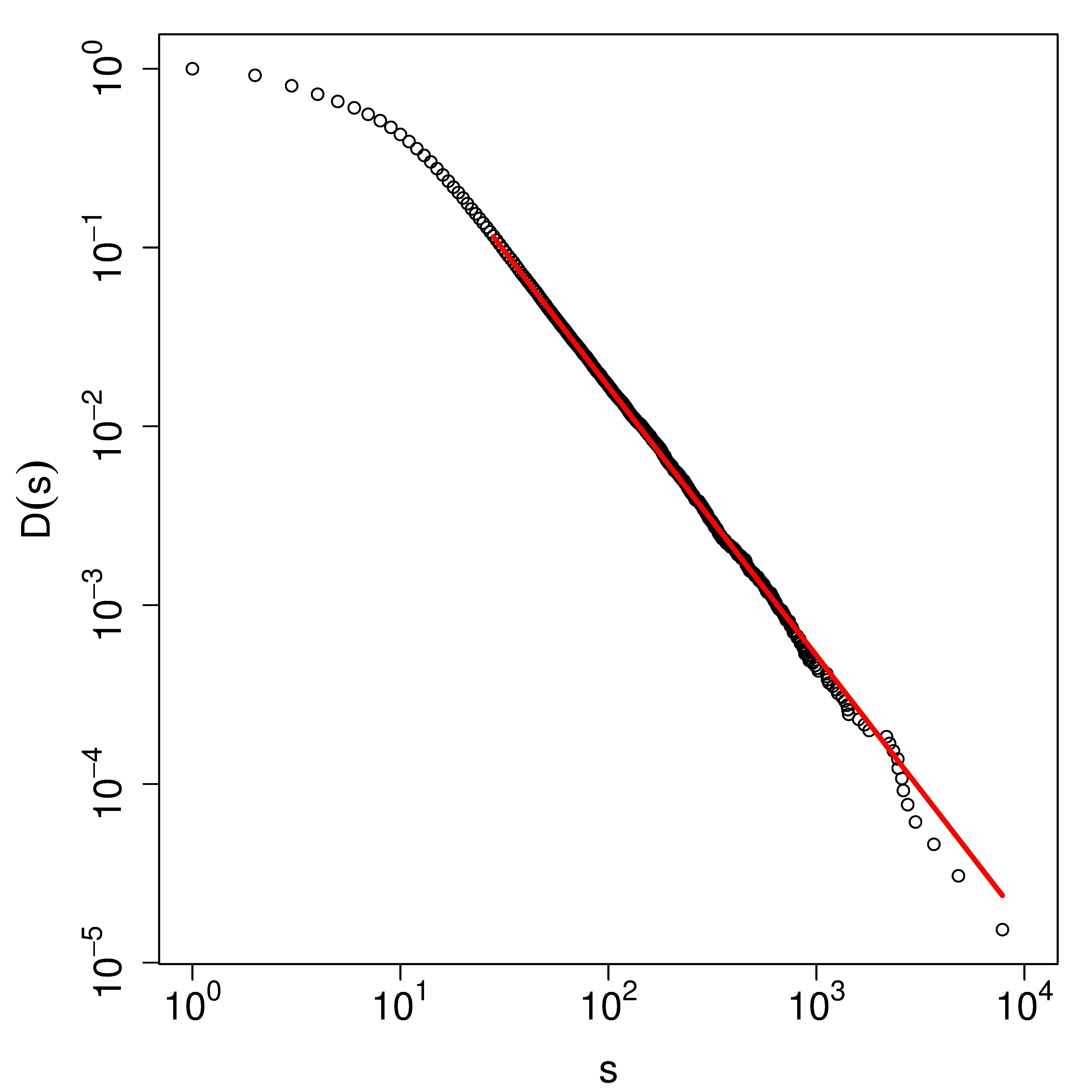}
\end{center}
\caption{(color online)
{\bf The complementary cumulative distribution function $D(s)$ of a community with size $s$ at the second modular level.}
A power-law fit to the data (red line) using the maximum likelihood estimation technique yields $D(s)\sim s^{-\gamma +1}$ 
with $\gamma =2.50 \pm 0.02$, $s_{min}=28.2 \pm 7.6$, and $p~value = 0.976$.}
\label{fig2}
\end{figure}

The cumulative distribution of the community size at this level is plotted
in Fig~\ref{fig2}.
We use maximum likelihood estimation (MLE)~\cite{clauset2009power} to quantitatively 
fit a statistically significant power-law
decay for the tail of the CCDF, which has the functional form $D(s) \sim s^{-\gamma+1}$ with
$\gamma=2.50\pm0.02$. The results indicate that the size of the communities is highly heterogeneous and spans 
over several orders of magnitude. 


We also analyzed the overexpressions of selected subcommunities. 
In terms of subcommunities, we observe wholesale and retail trade is the dominate overexpress attribute of
the five largest subcommunities of the largest community. The Kansai region is the only overexpressed region in the 2nd largest subcommunity of the largest community. In $C_{2:1}$, transport and postal activities, accommodations, eating and drinking services, living related and personal services, and amusement services dominate the overexpressed sectors, which are mainly based in urban regions (Tokyo and Chubu). The manufacturing, wholesale and retail trades in Tokyo and the Kansai region are overexpressed in $C_{2:2}$. 
Wholesale and retail trade dominate the overexpressed attribute in $C_{2:3}$, $C_{2:4}$ and $C_{2:5}$.
A detailed account of the results is provided in S1 Appendix.

\begin{figure}[!h]
\begin{center}
\includegraphics[width=0.5\textwidth]{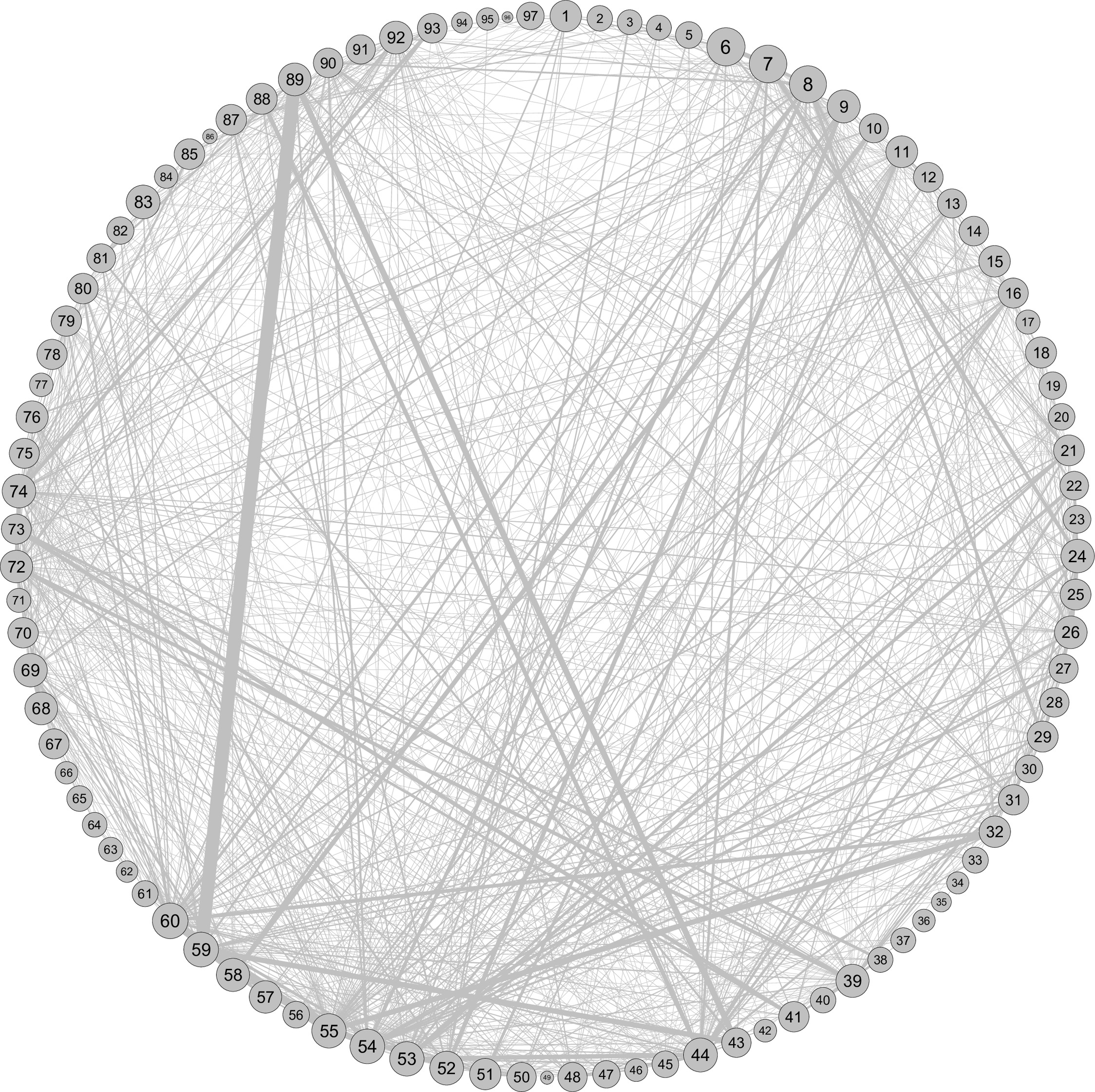}
\end{center}
\caption{{\bf Overexpression network of sectors.} The node size represents the percentage of firms belong to that particular sector.}  
\label{fig:oenet}
\end{figure}

\begin{figure}[!h]
\begin{center}
\includegraphics[width=0.7\textwidth]{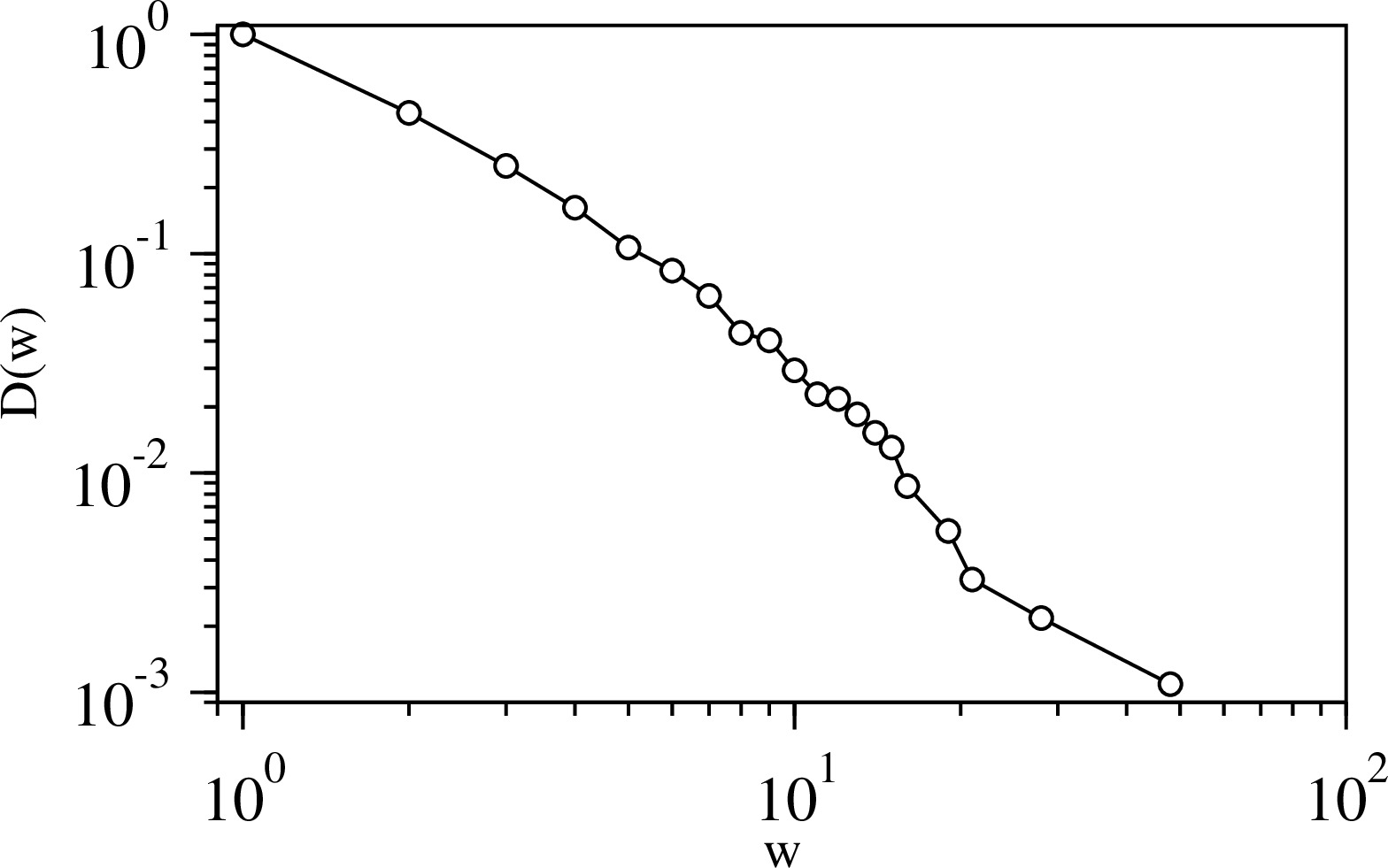}
\end{center}
\caption{\bf The complementary cumulative distribution of link-weight in the overexpression network.} 
\label{fig:link-weight}
\end{figure}

The network diagram in Fig~\ref{fig:oenet} shows the
overlapping nature of the industrial sectors in the communities.
We construct a weighted undirected network of 97 major sectors from sector over expression data for the
2nd modular level. Here, a weighted link of value 1 is formed between a pair of sectors
if they are overexpressed in the same community. The link-weight of the network is found to be highly
heterogeneous with a horizontal distribution as shown in Fig~\ref{fig:link-weight}. 
The top five heaviest weighted links between the sectors are listed in Table~\ref{tab:weight}. 

\begin{table}[ht]
\centering
\caption{\bf Top five heaviest weighted links between sectors:}
\begin{tabular}{|r|p{4cm}|p{4cm}|r|}
\hline
Rank & Node 1 & Node 2  & Weight\\ \hline
$1$  &Retail trade (machinery and equipment) &Automobile maintenance services  & 48 \\ 
$2$  & Miscellaneous wholesale trade  & Miscellaneous retail trade & 28 \\ 
$3$  &Road passenger transport & Automobile maintenance services & 21\\
$4$  &Miscellaneous manufacturing industries & Miscellaneous wholesale trade  & 19 \\ 
$5$  &Road passenger transport &Retail trade (machinery and equipment) &  19\\ \hline
\end{tabular}
\label{tab:weight}
\end{table}

\begin{figure}[ht]
\begin{center}
\includegraphics[width=0.7\textwidth]{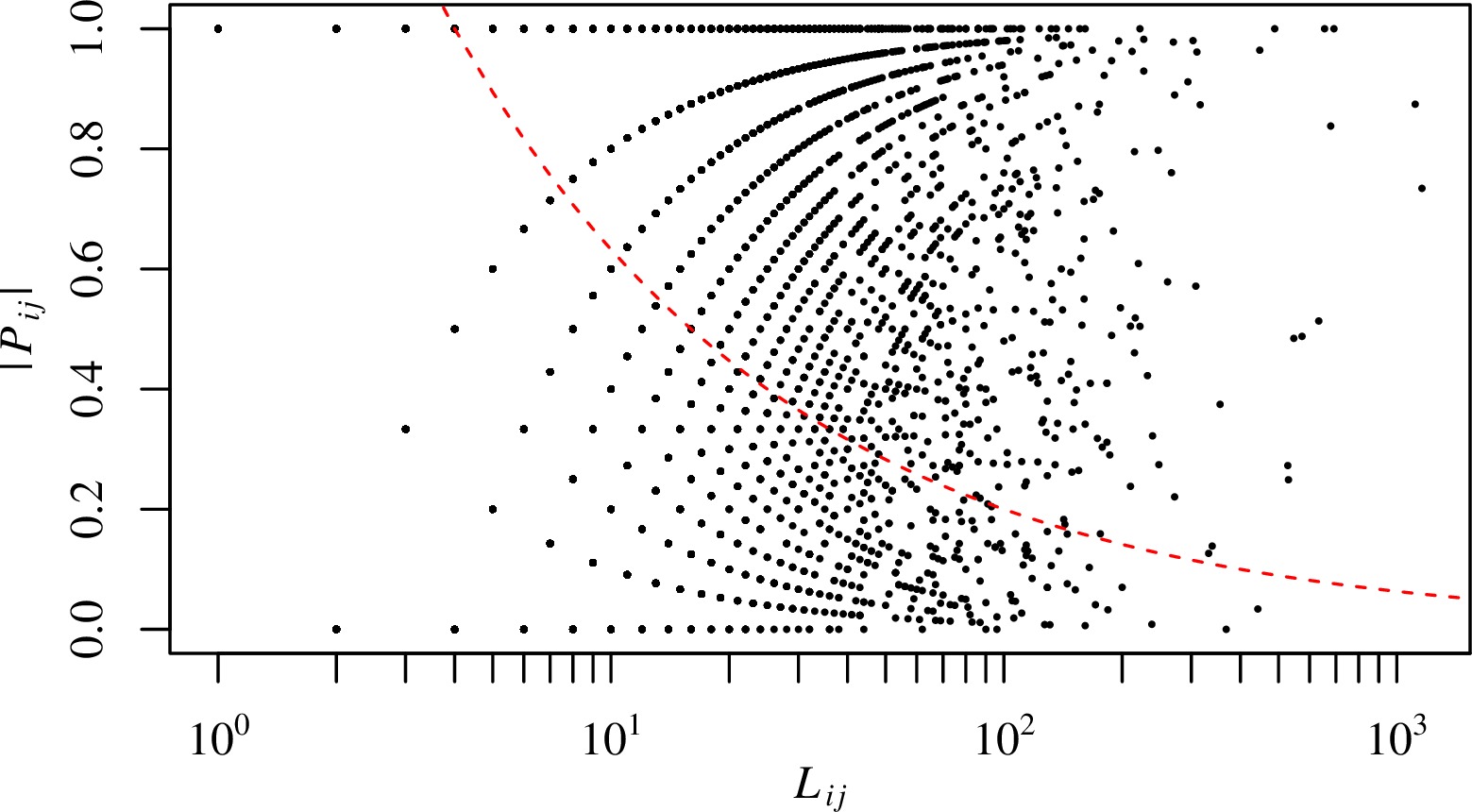}
\end{center}
\caption{{\bf Polarizability of the direction of the links interconnecting communities at the second level.}
Here, 1086 communities containing over 100 firms are selected. The dashed curve represents the same significance level as in Fig~\ref{fig:poltop}.}\label{fig:pol}
\end{figure}

Fig~\ref{fig:pol} is the same plot as Fig~\ref{fig:poltop}, but this new plot includes communities at the 2nd modular level. We can confirm that the links between the subcommunities are well polarized. Once again, this result is consistent with the nature of the map equation, which extracts communities of tightly connected nodes in a bidirectional way in a directed network.



\begin{figure}[ht]
\begin{center}
\includegraphics[width=0.75\textwidth]{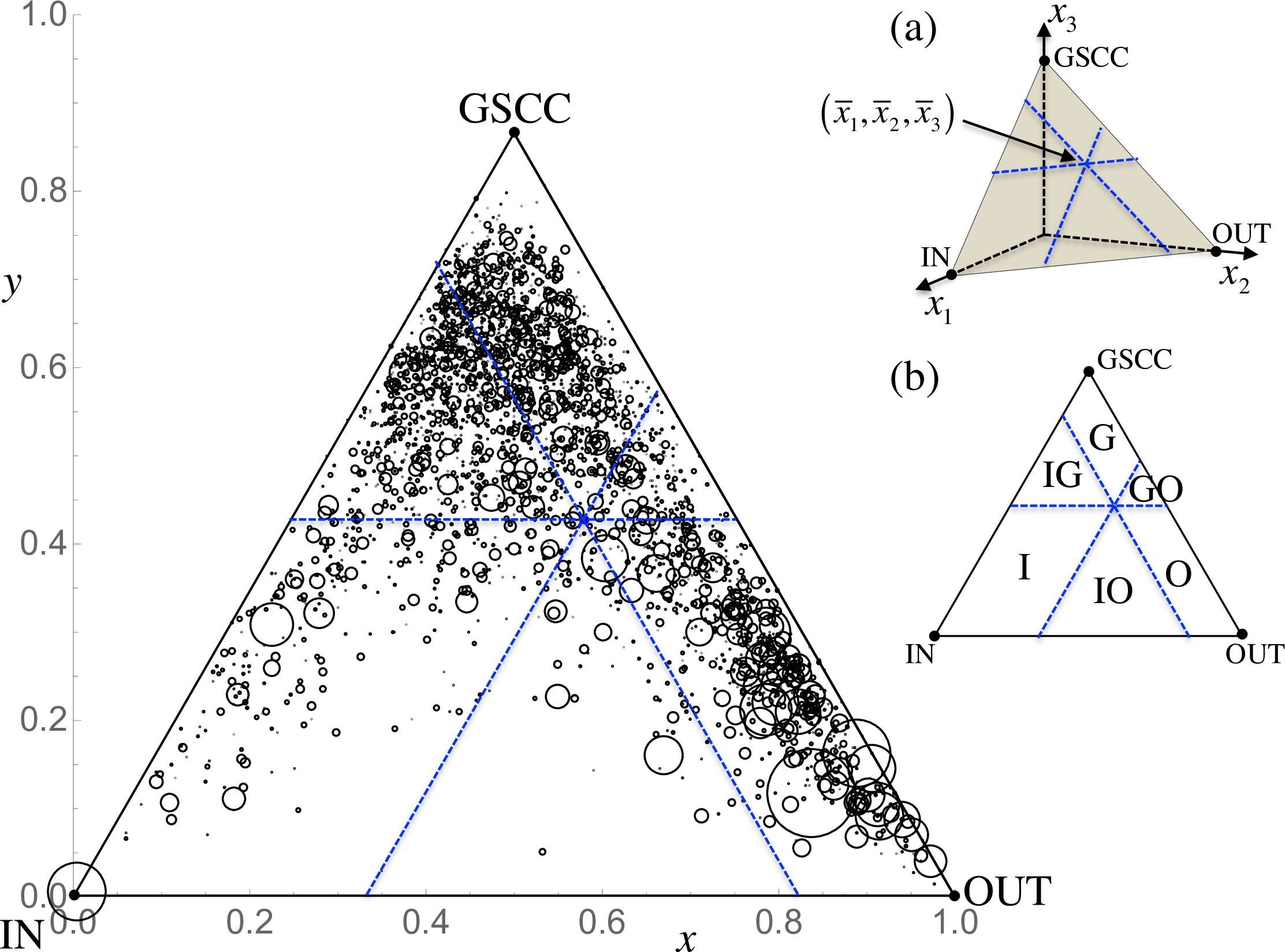}
\end{center}
\caption{{\bf Triangular diagram classifying communities at the second level by their relationship with the walnut structure.}
Each community is depicted by a circle located at point ($x$, $y$) inside the equilateral triangle, which corresponds to the composition ($x_{1}$, $x_{2}$, and $x_{3}$) of firms belonging to the IN, OUT, and GSCC components that are represented in three-dimensional space; the one-to-one correspondence between ($x$, $y$) and ($x_{1}$, $x_{2}$, $x_{3}$) is illustrated in the associated figure (a). The size of the communities is reflected by the area of their associated circles. The triangular region is decomposed into six domains with the average composition ($\bar{x}_{1}$, $\bar{x}_{2}$, $\bar{x}_{3}$) of the IN, OUT, and GSCC components for all firms, as designated in the associated figure (b); see the text for more detailed information on the domain decomposition.}\label{fig:triangular_diagram}
\end{figure}

Fig~\ref{fig:triangular_diagram} shows how mixed the IN, OUT, and GSCC components of the walnut structure are in each of the large communities with more than 50 firms at the 2nd level, adopting a triangular diagram representation.
We exclude firms belonging to TE; however, these are minor components of the walnut structure.
Here, 3,011 communities containing more than 50 firms are selected, for a total of 421,779 firms. Suppose that a community contains firms belonging to the IN, OUT, and GSCC components for which the percentages are given by $x_{1}$, $x_{2}$, and $x_{3}$, respectively. The walnut composition of the community is described by point ($x_{1}$, $x_{2}$, $x_{3}$) on the plane of $x_{1}+x_{2}+x_{3}=1$ in three-dimensional space. One can thereby establish one-to-one correspondence between a point inside an equilateral triangle and a composition of the three walnut components. 
The averaged composition of all the firms in the selected communities
(i.e., the total number of firms in the IN/OUT/GSCC components divided by the total number of firms in the selected communities)
is given by $\bar{x}_{1}=0.174$, $\bar{x}_{2}=0.333$, and $\bar{x}_{3}=0.493$. The triangular region in Fig~\ref{fig:triangular_diagram} is then decomposed into six domains in reference to $\bar{x}_{1}$, $\bar{x}_{2}$, and $\bar{x}_{3}$: the communities in domain G ($x_{1}<\bar{x}_{1}$, $x_{2}<\bar{x}_{2}$, $x_{3}>\bar{x}_{3}$) are GSCC-dominant; those in IG ($x_{1}>\bar{x}_{1}$, $x_{2}<\bar{x}_{2}$, $x_{3}>\bar{x}_{3}$) are GSCC-IN hybrid; those in I ($x_{1}>\bar{x}_{1}$, $x_{2}<\bar{x}_{2}$, $x_{3}<\bar{x}_{3}$) are IN-dominant; those in IO  ($x_{1}>\bar{x}_{1}$, $x_{2}>\bar{x}_{2}$, $x_{3}<\bar{x}_{3}$) are IN-OUT hybrids; those in O ($x_{1}<\bar{x}_{1}$, $x_{2}>\bar{x}_{2}$, $x_{3}<\bar{x}_{3}$) are OUT-dominant; and those in GO ($x_{1}<\bar{x}_{1}$, $x_{2}>\bar{x}_{2}$, $x_{3}>\bar{x}_{3}$) are GSCC-OUT hybrids. The total number of communities and firms in each domain are listed in Table~\ref{tab::triangular_diagram}. 
We observe that there are relatively fewer communities in the I domain and more communities in the IG domain.
The IN components thus tend to combine with the GSCC components to form a single community.
On the other hand, 
there are an appreciable number of communities dominated by the OUT components, leading to relatively few communities of IN-OUT and GSCC-OUT hybrids.
This tendency, in terms of the characteristics of the communities, may reflect the industrial structure of Japan, which imports raw materials and produces a wide variety of goods out of these 
for both export and domestic consumption. We are also interested in what occurs in other countries. Once data on the production networks of other countries is available, we hope to compare their community characteristics with those of Japan.

\begin{table}[ht]
\centering
\caption{\bf Classification of communities at the second level based on the walnut structure}\label{tab::triangular_diagram}
\begin{tabular}{|c|r|r|}
\hline
Domain & \#com & \#firms\\ \hline
G      & 1,010  & 114,399\\ 
IG      &  841  & 92,163\\
I      &  294  & 44,563\\
IO      &   80  & 14,362\\
O      &  640  & 139,986\\
GO      &  146  & 16,306\\ \hline
Total  & 3,011 & 421,779\\ \hline
\end{tabular}
\begin{flushleft}
``\#com" and ``\#firms" refer to the total number of communities and firms, respectively, in each of the six domains defined in Fig~\ref{fig:triangular_diagram}(b).
\end{flushleft}
\end{table}

Although the IN components tend to to merge with the GSCC, we can see the large circle at the vertex of Fig~\ref{fig:triangular_diagram}. On the other hand, Table \ref{tab:walnut_dist} shows that most nodes in the IN component have a distance of 1 from the GSCC. Therefore, one may think that there is a large community almost purely composed of nodes in the IN components of the Walnut shape (Fig~\ref{fig:walnut2}). Actually, this configuration indicates an interesting structure where the nodes are mutually connected and simultaneously connected to nodes in the GSCC. It can be precisely said that the  community is in the shape of a walnut shell.


\section*{Comparison of industrial sectors}
\label{cha:density}

As is mentioned in the Introduction Section,
detecting communities in the supply-chain network
is crucial for understanding the
agglomerative behavior of firms.
This type of research is important because the detected communities are densely connected,
and it is plausible that these firms affect each other
through the links.

On the other hand, industrial sectors commonly label firms,
and these labels are widely used in the economics literature.
If there is no difference
between the detected communities and the industrial sectors,
then there is no reason to make an effort to detect these communities.
Therefore, in this section,
we show how the detected communities are different
from industrial sectors in terms of the interconnections
between the groups.

Although different classifications are used for industrial sectors,
we discuss the one used in the input-output table \cite{Leontief36}.
We use this classification because the input-output table is a major research domain in economics, and, more importantly,
the purpose of the input-output table is to discuss money flows,
which corresponds to the purpose of this paper.

As previously mentioned, there are 209 communities in the 1st level and 66,133 communities in the 2nd level.
On the other hand, the input-output tables have 13, 37, 108, 190, and 397 sectoral classifications, which are nested.
We choose to compare 209 communities and 190 industrial sectors because these numbers are comparable.

\begin{figure}[!h]
\begin{center}
\includegraphics[width=0.8\textwidth]{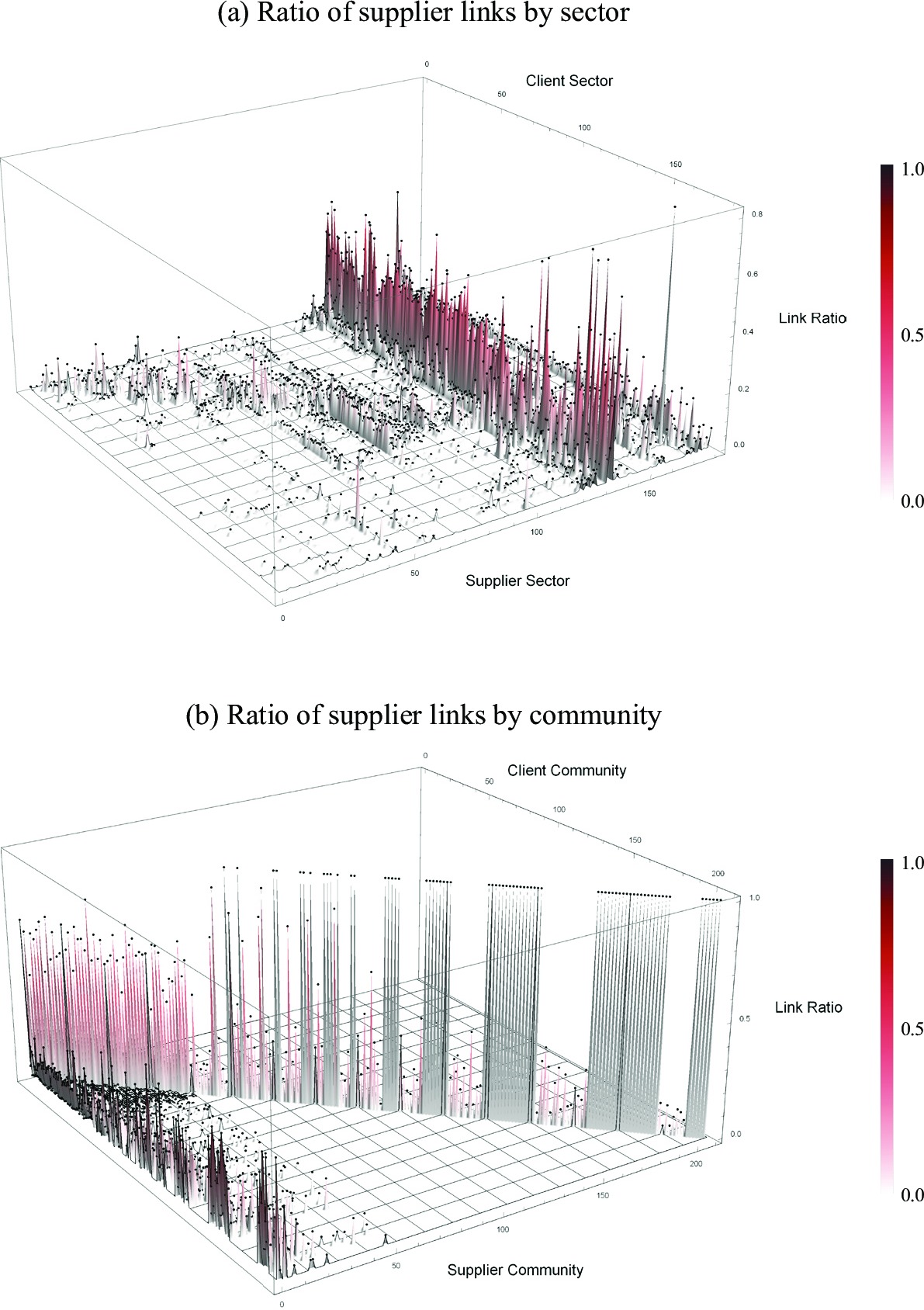}
\label{fig:kseccom}
\end{center}
 \caption{{\bf Density of links over intergroups}
These figures show how many links the intergroups have.
The top figure (a) shows the 3D plots of the industrial sectors.
The bottom figure (b) shows the 3D plots of the communities.}
 \label{fig:dense}
\end{figure}

First, we counted the number of links between the communities and the industrial sectors.
Fig~\ref{fig:dense} shows the difference between these two groups.
These figures correspond to matrices that show the number of links in row groups and column groups.
Each element is divided by the sum of its row.

If the intra-links within the groups are dominant, then the diagonal elements of these matrices should have high density.
As is shown in Fig~\ref{fig:dense}, we can find the diagonal elements because the communities are denser than the other elements.
However, the diagonal elements of the sectors do not have dense links.
We see a vertical line in the matrix instead.
The suppliers in the line include 5111: Wholesale and 5112: Retailing,
and this result is natural because firms sell their products to industrial sectors.
The overall ratio of intra-links, i.e., (the number of intra-group links)/(the number of all links) is
20.9\% for industrial sectors and 63.3\% for communities.

We can conclude that the detected communities in this paper explicitly illustrate the agglomeration of firms based on supply-chain networks
rather than industrial sectors, which is more commonly used to categorize firms.
This result also tells us that communities with densely connected firms consist of various industrial sectors, and they have their own economies,
i.e., small universes.

In this paper, we do not weight the links of the network. However, obviously, each transaction has a value, and there is a diversity of transactions.  We can estimate the weights by using the sales of the firms. If we have totally different results with the results we have obtained here, a further analysis might be necessary. However, the additional analyses based on weighting the links in the networks do not show any significant difference. The details of these results are shown in S1 Appendix: Intra-link density of the weighted links.

\section*{Conclusion and discussion}

We analyze the overall structure and hierarchical communities embedded 
in the production network of one million firms and five million links that represent trade relationships in Japan in 2016, with the aim of 
simulating the macro/micro level dynamics of the economy.

For the former, we find that the IN and OUT components (20 \% and 26\% of the firms)
form tight shells (semi-spheres) around the GSCC component, which we call a ``walnut" structure rather than a ``bow-tie" structure, which is well-known for representing web networks and other type of networks that have loose wings made of IN and OUT components.

For the latter, we use the Infomap method to detect a hierarchy that includes 5 layers of communities, of which
most of the irreducible (those that do not have any lower level subcommunities) belong to the 2nd level.
Furthermore, the size distribution of the 2nd level communities show clear power-law behavior at the large end.
In addition to the large number of irreducible communities made primarily of GSCC components and those that exist in IN shells or Out shells, there is a fair number of communities made of IN and GSCC components, GSCC and OUT components, and even IN and OUT components. These communities are expected due to the walnut shape of the overall structure: IN and OUT components are not far from each other as they are in the bow-tie structure, but they form tight shells, whose ends are closely woven with each other.
Furthermore, we examine the overexpression of the major communities in terms of industrial sectors and prefectures and find that they are not formed within a sector but span several sectors and prefectures. 
These communities have various shapes:
in some cases, they are formed around goods and services related to a particular item, such as food. Sometimes these communities are made of small firms connected with a major hub such as a large construction company in a particular prefecture or a medical insurance agency.

These findings have major implications for the study of the macro economy: Consider an economic crisis. Once this crisis starts, whether it is due to a natural disaster in a particular region of a country or a major failure of a large company, it is expected that it initially affects the community in which this region or company is located. 
Then the effects of this crisis will spread to other neighboring communities.
This analysis is very different from input-output analysis and is expected to be useful because an input-output analysis is based on the assumption that firms in the same sectors are well-connected with each other.
In contrast, what we find is that the effects of a crisis will spread throughout communities rather than industries.
The hierarchical community structure studied in this paper can be immediately applied to the analysis of large-scale modelling and simulation: the macro economy of a country or countries is an aggregation of products that economically affect the trade network as well as a multitude of networks of networks. Constructing models that span all the networks would be an interesting but exhaustive elaboration of this work. Instead, we may study one community at a time and then connect the results to obtain an overall picture.
Research in this direction has already begun and will appear in the near future (\cite{krichene2017business, Hazem, arata}).


\section*{Acknowledgments}
We are grateful to Y. Ikeda, W. Souma and H. Yoshikawa for their insightful comments
and encouragement.
We are also grateful to Tokyo Shoko Research Ltd.\ and RIETI for making this research
possible by providing us with the production network data.
This study was supported in part by the Project “Large-scale Simulation and Analysis of Economic Network for Macro Prudential Policy” undertaken at the Research Institute of Economy, Trade and Industry (RIETI), the Ministry of Education, Culture, Sports, Science and Technology (MEXT) as Exploratory Challenges on Post-K computer (Studies of Multi-level Spatiotemporal Simulation of Socioeconomic Phenomena), the Grant-in-Aid for Scientific Research (KAKENHI) by JSPS grant numbers 25400393 and 17H02041 and the Kyoto University Supporting Program for Interaction-based Initiative Team Studies: SPIRITS, as a part of the Program for Promoting the Enhancement of Research Universities, MEXT, JAPAN. There was no additional external funding received for this study.



\section*{Supporting information}

\paragraph*{S1 Appendix.}
\label{S1_Appendix}
{\bf Appendix to the manuscript.}

\clearpage
\clearpage
\clearpage
\clearpage

\newcommand{\beginsupplement}{%
        \setcounter{table}{0}
        \renewcommand{\thetable}{\Alph{table}}
        \setcounter{figure}{0}
        \renewcommand{\thefigure}{\Alph{figure}}
     }
\newcommand{\eq}[1]{Eq.~(\ref{#1})}
\newcommand{\fig}[1]{Fig.~\ref{#1}}

\newcommand{\dts}{\partial_t}
\newcommand{\dxs}{\partial_x}
\newcommand{\dvs}{\partial_v}


\beginsupplement
\title{Appendix S1: Hierarchical communities in the walnut structure of the Japanese production network}


\maketitle
\section{Data classifications}

Table \ref{tab:numfirm_division} lists the number of firms in 20
industrial sectors.

\begin{table}[htbp]
\centering
\caption{\bf Industrial sectors and firm distribution}
\begin{tabular}{|r|c|l|r|r|}
\hline
ID & Code & Sector & \# Firms & \% \\
\hline
1 & A & Agriculture & 9,841 & 0.92 \\
2 & B & Fisheries & 1,211 & 0.11 \\
3 & C & Mining & 1,268 & 0.12 \\
4 & D & Construction & 357,199 & 33.51 \\
5 & E & Manufacturing & 156,188 & 14.65 \\
6 & F & Electricity, Gas, Heat Supply \& Water ({\it EGW)} & 1,470 & 0.14 \\
7 & G & {\it Information} \& Communications & 26,539 & 2.49 \\
8 & H & {\it Transport} \& Postal & 36,736 & 3.45 \\
9 & I & Wholesale \& {\it Retail} Trade& 254,251 & 23.85 \\
10 & J & {\it Finance} \& Insurance & 7,506 & 0.70 \\
11 & K & Real Estate & 41,837 & 3.92 \\
12 & L & {\it Scientific Research}, Professional \& Technical Services & 42,030 & 3.94 \\
13 & M & {\it Accommodations}, Eating/Drinking Services & 17,322 & 1.62 \\
14 & N & {\it Living-related}/Personal \& Amusement Services & 17,365 & 1.63 \\
15 & O & {\it Education}, Learning Support & 4,655 & 0.44 \\
16 & P & {\it Medical}, Health Care \& Welfare & 30,154 & 2.83 \\
17 & Q & Compound Services & 6,472 & 0.61 \\
18 & R & Other Services & 52,190 & 4.90 \\
19 & S & Government & 1,803 & 0.17 \\
20 & T & Unable to classify & 0 & 0.0 \\
\hline
\end{tabular}
\begin{flushleft}
The number of firms classified by
  industrial sectors, which is based on the Japan Standard Industrial Classification.
  The words in italics are abbreviated in the main text.
  \end{flushleft}
\label{tab:numfirm_division}
\end{table}

\begin{figure}
    \centering
    \includegraphics[width=0.5\textwidth]{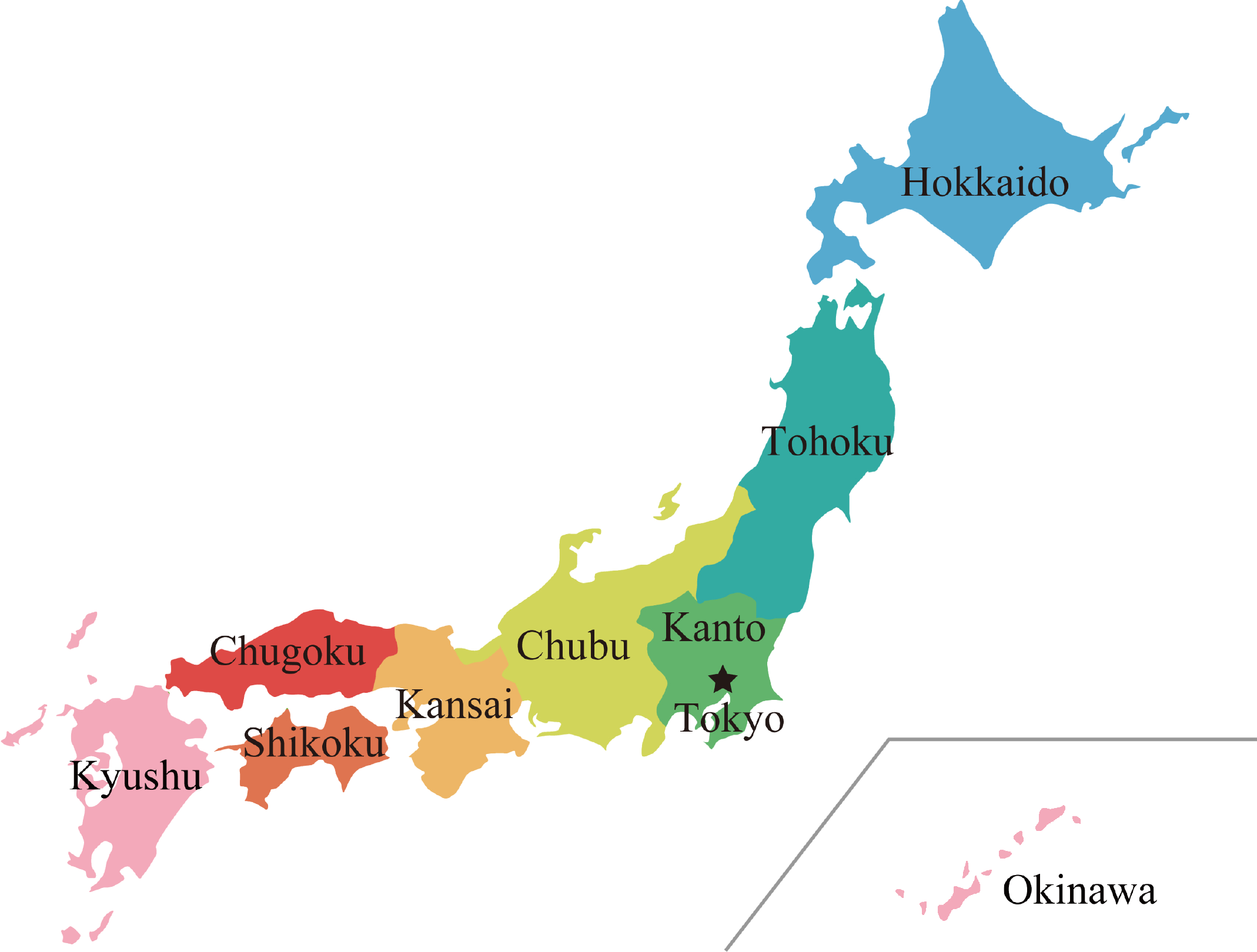}
    \caption{{\bf Eight regions and Tokyo, Japan.}}
    \label{fig:japan}
\end{figure}

Table~\ref{tab:numfirm_region} lists the number of firms in 8 regions 
and city of Tokyo, Japan,
which are illustrated in Figure \ref{fig:japan}.

\begin{table}[htbp]
\centering
\caption{\bf Regional areas and firm distribution}
\begin{tabular}{|r|l|r|r|}
\hline
id & region & \#firms & \% \\
\hline
1 & Hokkaido & 54,423 & 5.11 \\
2 & Tohoku & 87,374 & 8.20 \\
3 & Kanto & 187,186 & 17.56 \\
4 & Tokyo & 146,614 & 13.75 \\
5 & Chubu & 196,477 & 18.43 \\
6 & Kansai & 168,701 & 15.83 \\
7 & Chugoku & 69,312 & 6.50 \\
8 & Shikoku & 40,397 & 3.79 \\
9 & Kyusyu-Okinawa & 115,553 & 10.84 \\
\hline
\end{tabular}
\begin{flushleft}
The number of firms in each regional area is determined by the
  geographical location of the main office of the firm.
  ``Kanto" means ``Kanto less Tokyo", as greater ``Tokyo" is in the ``Kanto" region.
  \end{flushleft}
\label{tab:numfirm_region}
\end{table}

\clearpage
\section{Comparison of the community structure based on modularity maximization}\label{som:comp}

\begin{figure}[!ht]
\begin{center}
\includegraphics[width=0.7\textwidth]{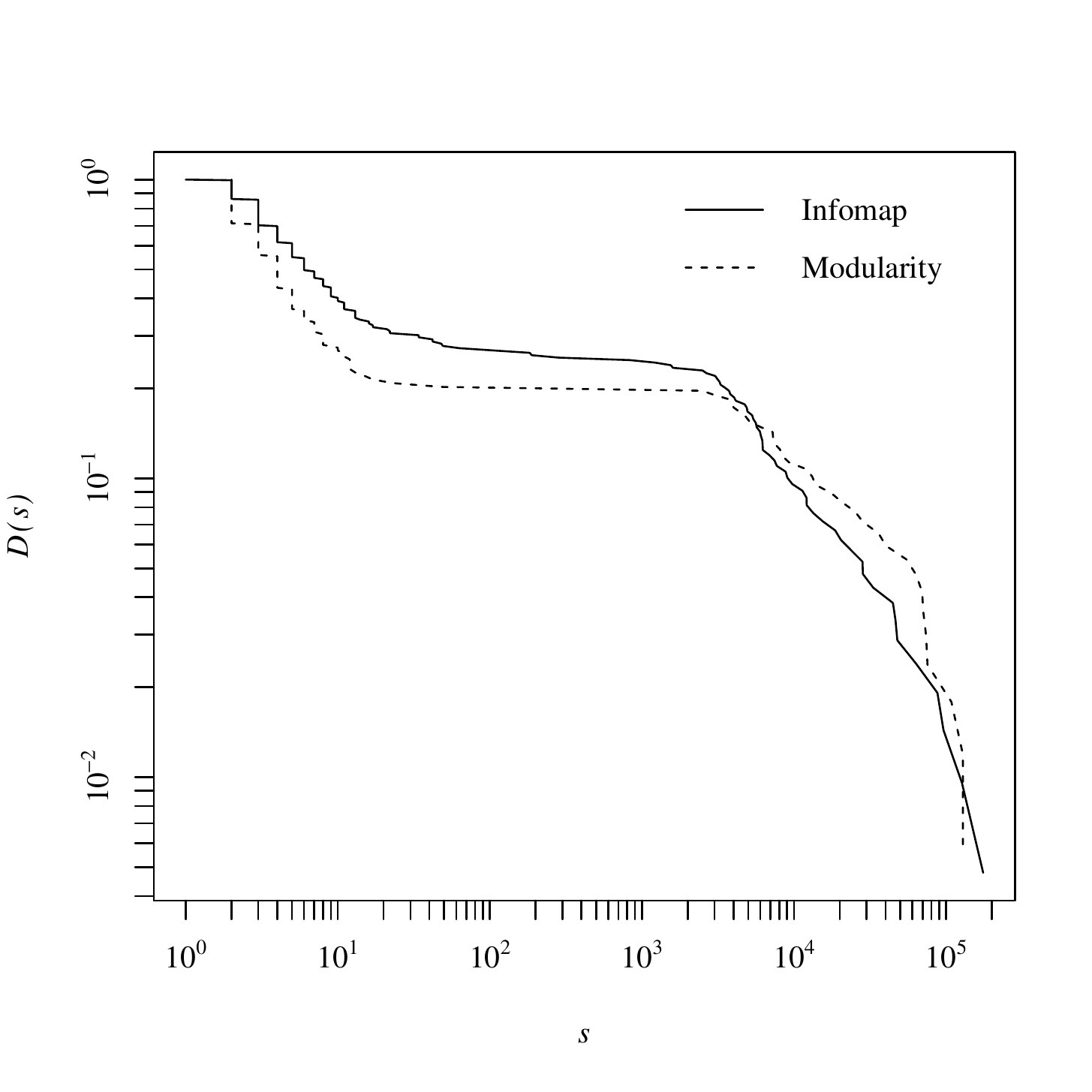}
\end{center}
\caption{
{\bf The complementary cumulative distribution function $D(s)$ of communities with size $s$ at the top modular level.}
}
\label{figB}
\end{figure}

Fig~\ref{figB} shows the complementary cumulative distribution of the community size at the top level, which is compared with the corresponding result obtained by the modularity maximization method.
The two distributions are quite similar, indicating that the two community structures are similar. The distribution for modularity maximization is obtained for a undirected network, 
that is, by ignoring the direction of the links.
We also conducted directed modularity analysis, but the result does not differ much from those reported above. 

One can quantify the similarity between the 1st level of the community structure obtained with the map equation and that obtained with modularity in terms of the Rand index~\cite{doi:10.1080/01621459.1971.10482356}, which is a measure of similarity between two data clusterings. The adjusted Rand index, in which the coincidental chance that a clustering of a pair of nodes is identical, is subtracted from the original index and calculated as 0.325. This large value indicates that the two partitions resemble each other significantly. 

In Table~\ref{tab:jaccard}, a more detailed comparison between the two community structures is made, that is, a community by community comparison is made using the Jaccard index~\cite{Jaccard1901}, which measures the similarity between two sets. We see that there is remarkable one-to-one correspondence between the two partitions. The major communities of the map equation, up to the 7th largest in Table~\ref{tab:jaccard}, have counterparts in the partition with modularity.

\begin{table}[ht]
\caption{\bf Jaccard index between the major communities of the top level obtained with modularity and those obtained with the hierarchical map equation}\label{tab:jaccard}
\centering
\begin{tabular}{r|rrrrrrrrrr}
  \hline
 & 1 & 2 & 3 & 4 & 5 & 6 & 7 & 8 & 9 & 10 \\ 
  \hline
1 & 0.02 & 0.64 & 0.01 & 0.01 & 0.01 & 0.01 & 0.01 & 0.02 & 0.01 & 0.01 \\ 
  2 & 0.41 & 0.01 & 0.01 & 0.03 & 0.01 & 0.00 & 0.00 & 0.05 & 0.01 & 0.01 \\ 
  3 & 0.04 & 0.01 & 0.31 & 0.01 & 0.01 & 0.00 & 0.00 & 0.03 & 0.00 & 0.00 \\ 
  4 & 0.02 & 0.03 & 0.01 & 0.01 & 0.11 & 0.09 & 0.01 & 0.12 & 0.01 & 0.20 \\ 
  5 & 0.03 & 0.00 & 0.14 & 0.01 & 0.00 & 0.00 & 0.00 & 0.02 & 0.00 & 0.00 \\ 
  6 & 0.03 & 0.03 & 0.01 & 0.28 & 0.01 & 0.00 & 0.01 & 0.03 & 0.04 & 0.02 \\ 
  7 & 0.05 & 0.01 & 0.02 & 0.02 & 0.32 & 0.01 & 0.01 & 0.01 & 0.01 & 0.03 \\ 
  8 & 0.12 & 0.01 & 0.05 & 0.01 & 0.02 & 0.00 & 0.00 & 0.02 & 0.02 & 0.00 \\ 
  9 & 0.01 & 0.03 & 0.01 & 0.01 & 0.01 & 0.01 & 0.50 & 0.02 & 0.00 & 0.01 \\ 
  10 & 0.01 & 0.01 & 0.00 & 0.00 & 0.01 & 0.39 & 0.06 & 0.01 & 0.01 & 0.00 \\ 
   \hline
\end{tabular}
\begin{flushleft}
The 10 largest communities of the map equation are aligned in the horizontal direction and those of modularity, in the vertical direction. Note that the Jaccard index takes $1/3$ for two sets of equal size, and 50\% of the elements overlap.
\end{flushleft}
\end{table}



\clearpage
\section{Overexpressions: level 1}
In the main text, we present the overexpression of 20 sector divisions and 9 regions in the selected large communities.
Here, we further expose the overexpression of 99 sectors and 47 prefectures in the following communities. 

Rank: 1

Over-expression of prefectures: Ibaraki; Gunma; Saitama; Chiba; Tokyo; Kanagawa; 
Nagano; Shizuoka; Aichi; Mie; Osaka; Hyogo; and Hiroshima

Over-expression of sectors: Equipment installation work; Manufacture of plastic products, except otherwise classified; 
Manufacture of rubber products; Manufacture of iron and steel;  Manufacture of non-ferrous metals and products; Manufacture of
fabricated metal products; Manufacture of general purpose machinery; Manufacture of production machinery; 
Manufacture of business oriented machinery;  Electronic parts, devices and electronic circuits;
Manufacture of electrical machinery, equipment and supplies; Manufacture of information and communication electronics equipment;
Manufacture of transportation equipment; Production, transmission and distribution of electricity; Heat supply; Wholesale
trade (general merchandise); Wholesale trade (building materials, minerals and
metals, etc); Wholesale trade (machinery and equipment); Retail trade (machinery and equipment); Machine, etc.; and repair services, except those that are otherwise classified

Rank: 2

Overexpression of prefectures: Hokkaido; Aomori; Iwate; Miyagi; Akita; Yamagata; Fukushima; Niigata;
Yamanashi; Nagano; Shizuoka; Tottori; Shimane; Tokushima; Kagawa; Ehime; Kochi; Saga; Nagasaki; Kumamoto; 
Oita; Miyazaki; and Kagoshima

Overexpression of sectors: Agriculture; Fisheries, except Aquaculture; Aquaculture; Manufacture
of food; Manufacture of beverages, tobacco and feed; Railway transport; Warehousing; Wholesale
trade (general merchandise); Wholesale trade (food and beverages); Miscellaneous wholesale trade;
Retail trade (general merchandise); Retail trade (food and beverage); Non-store retailers; 
Financial institutions for cooperative organizations; Non-deposit money corporations, including
lending and credit card businesses; Real estate lessors and managers; Accommodations; Eating and
drinking places; Food take out and delivery services; Miscellaneous living-related and personal services; 
Services for amusement and recreation; Social insurance, social welfare and care services; 
Cooperative associations and not elsewhere classified; and Miscellaneous services

Rank:3 

Overexpression of prefectures: Ibaraki; Saitama; Chiba; Tokyo; Kanagawa; and Osaka

Overexpression of sectors: Mining and quarrying of stone; Construction work, general including public
and private construction work; Construction work by specialist contractors, except equipment installation
work; Equipment installation work; Manufacture of ceramic, stone and clay products; Collection, purification
and distribution of water and sewage collection, processing and disposal; Railway transport; Financial
auxiliaries; Real estate agencies; Real estate lessors and managers; Goods rental and leasing; Technical
services and not elsewhere classified; Social insurance, social welfare and care services; Automobile maintenance
services; Political, business and cultural organizations; and Local government services

Rank: 4

Overexpression of prefectures: Aomori; Iwate; Miyagi; Akita; Fukushima; Ibaraki; Tochigi; Gunma;
Saitama; Chiba; Kanagawa; Ishikawa; Shizuoka; Aichi; Mie; and Saga

Overexpression of sectors: Manufacture of petroleum and coal products; Manufacture of transportation
equipment; Road Passenger transport; Road freight transport; Water transport; Warehousing; Services
incidental to transport; Wholesale trade (machinery and equipment); Retail trade (machinery
and equipment); Miscellaneous retail trade; Insurance institutions, including insurance agents brokers
and services; Goods rental and leasing; School education; Waste disposal business; and Automobile maintenance
services

Rank: 5

Over-expression of prefectures: Tokyo; Kanagawa; and Osaka

Over-expression of sectors: Printing and allied industries; Manufacture of business oriented machinery;
Electronic parts, devices and electronic circuits; Manufacture of information and communication electronics
equipment; Miscellaneous manufacturing industries; Communications; Brodcasting; Information services; 
Services incidental to the internet; Video picture information, sound information, and character information
production and distribution; Air transport; Wholesale trade (machinery and equipment); Retail trade
(machinery and equipment); Miscellaneous retail trade; Non-store retailers; Banking; Financial institutions
for cooperative organizations; Non-deposit money corporations, including lending and credit card businesses;
Financial products transaction dealers and futures commodity dealers; Financial auxiliaries; Real estate
lessors and managers; Goods rental and leasing; Professional services and not elsewhere classified; Advertising; Technical
services and not elsewhere classified; Miscellaneous living-related and personal services; Services for amusement and recreation;
School education; Miscellaneous education and learning support; Employment and worker dispatching services;
Miscellaneous business services; and Political, business and cultural organizations

\section{Overexpressions: Second modular level}

A short summary of the overexpression of 20 sector divisions and 9 regions in the 2nd level communities is provided 
in Tables \ref{table_2nd_modular_A} and \ref{table_2nd_modular_B}.

\begin{table}[ht]
\small
\centering
\caption{\bf Overexpressions of the 2nd level communities}
\scalebox{0.75}{
\begin{tabular}{|c|r|r|p{4.0cm}|p{5.0cm}|ccc|}
  \hline
Index & Size & Rank &Region & Sector & IN & GSCC & OUT\\ 
  \hline
  1:1 & 2,618 & 6 &Hokkaido (0.04); Tohoku (0.10); Chugoku (0.09); Shikoku (0.05); Kyusyu-Okinawa (0.12) & Retail (0.85) & 0.03 & 0.10 & 0.86\\ \hline
  1:2 & 1,430 & 16 &Kansai (0.32) & Transport (0.08); Retail (0.31)& 0.28 & 0.68 & 0.03\\ \hline
  1:3 & 1,132 & 25 &Hokkaido (0.04); Tohoku (0.12); Chugoku (0.11); Shikoku (0.05); Kyusyu-Okinawa (0.11) &  Retail (0.77) & 0.06 & 0.14 & 0.79\\ \hline
  1:4 & 863 & 38 &Hokkaido (0.06); Tohoku (0.11); Shikoku (0.04) & Retail (0.82) & 0.05 & 0.12 & 0.83
  \\ \hline 
  1:5 & 854 & 39  &Hokkaido (0.04); Chugoku (0.10); Kyusyu-Okinawa (0.13) & Retail (0.43); Accommodations (0.03) & 0.11 & 0.59 & 0.29\\ 
  \hline\hline
 2:1 & 2,474 &  8  &Tokyo (0.18); Chubu (0.23) & Information (0.02); Transport (0.13);  Accommodations (0.19); 
 Living-related (0.49); Education (0.009); Services N.E.C. (0.04) & 0.17 & 0.44 & 0.38\\ \hline
 2:2 & 1,200 & 23  & Tokyo (0.45); Kansai (0.22) &  Manufacturing (0.23); Retail (0.55); Scientific research (0.04) & 0.53 & 0.37 & 0.09\\ \hline
 2:3 & 1,121 & 27  & Kanto (0.20); Chubu (0.24); Kansai (0.18) & Manufacturing (0.26); Retail (0.60)  & 0.09 & 0.56 & 0.35\\ \hline
 2:4 & 1,022 & 28 &Kanto (0.22); Tokyo (0.25), Kansai (0.26) &  Retail (0.72)   & 0.12 & 0.34 & 0.53\\ \hline
 2:5 & 1,010 & 29  & Kanto (0.33) &Agriculture (0.13); Retail (0.67) & 0.10 & 0.49 & 0.40\\ \hline\hline
 3:1 & 2,173 & 12  & Hokkaido (0.02); Tohoku (0.05); Tokyo (0.33); Chubu (0.09); Chugoku (0.04); Shikoku (0.02); Kyusyu-Okinawa (0.05) & Manufacturing (0.05); Scientific research (0.09),
 Services N.E.C. (0.06)& 0.60 & 0.36 & 0.05\\ \hline
 3:2 & 834 & 41  &Tokyo (0.55) & Informations (0.02);  Real estate(0.73); 
  Scientific research (0.09)& 0.30 & 0.25 & 0.40\\ \hline
 3:3 & 776 & 46  &Tokyo (0.97) & Scientific research(0.12); Medical (0.32); Services, N.E.C. (0.22);
  Government (0.02) & 0.70 & 0.12 & 0.11\\ \hline
 3:4 & 740 & 53  &Tohoku (0.08); Chubu (0.16); Chugoku (0.06); Shikoku (0.02); Kyusyu-Okinawa (0.09)  & Construction (0.78) & 0.68 & 0.26 & 0.06\\  \hline
 3:5 & 547 & 87  &Hokkaido (0.02); Tohoku(0.06); Chubu (0.20); Chugoku (0.08); Shikoku (0.04); Kyusyu-Okinawa (0.12) & Construction (0.84) & 0.83 & 0.12 & 0.05\\ \hline
 \end{tabular}
}
\begin{flushleft} 
The overexpression in terms of regions and sector-divisions for the five largest communities at the 2nd level.
``Rank" refers to the rank among all the 2nd level subcommunities.
The percentage of nodes having a particular attribute is indicated in parentheses.
\end{flushleft}
\label{table_2nd_modular_A}
\end{table}

\begin{table}[ht]
\small
\centering
\caption{\bf Overexpressions of the 2nd level communities, continued}
\scalebox{0.75}{
\begin{tabular}{|c|r|r|p{4.0cm}|p{5.0cm}|ccc|}
  \hline
Index & Size&Rank &Region & Sector & IN & GSCC & OUT\\ 
  \hline
4:1 & 7,843 & 1  &Hokkaido (0.06);  Tokyo (0.20); Shikoku(0.04) &Information (0.02); Finance (0.54); Real estate (0.12);
 Scientific research (0.03) & 0.09 & 0.13 & 0.77 \\ \hline
  4:2 & 2,747 & 5 &Hokkaido (0.05); Kanto (0.27); Chubu (0.23) & Retail (0.87) & 0.04 & 0.34 & 0.61\\ \hline 
  4:3 & 2,249 & 11  &Shikoku (0.05); Kyusyu-Okinawa (0.13) & Retail (0.87) & 0.03 & 0.12 & 0.85\\ \hline 
 4:4 & 1,416 & 17 & Hokkaido (0.06); Shikoku (0.04) & Retail (0.88) & 0.03 & 0.13 & 0.83\\ 
 4:5 & 1,149 & 24  &Kansai (0.96) & Retail(0.46); Services, N.E.C. (0.36) & 0.06 & 0.43 & 0.51\\ 
 \hline\hline
  5:1 & 2,996 & 4 &Hokkaido (0.05); Tohoku (0.07); Kanto (0.17); Chubu (0.20); Kansai (0.16); Chugoku (0.06); Kyusyu-Okinawa (0.10)  & Living-related (0.54) & 0.09 & 0.23 & 0.62\\ \hline
  5:2 & 1,248 & 22  & Hokkaido (0.05); Tohoku (0.07); Kanto (0.22); Chubu (0.18) & Transport(0.04); Retail(0.62) & 0.11 & 0.22 & 0.66\\ \hline
  5:3 & 1,127 & 26 & Chubu (0.28) & Manufacturing(0.12); Retail (0.62), Education (0.09) & 0.09 & 0.29 & 0.61\\ \hline
  5:4 & 832 & 42  &Kansai (0.17) &Retail (0.33); Real estate (0.10); Living-related(0.27) & 0.16 & 0.39 & 0.43\\ \hline
  5:5 & 750 & 50  &Hokkaido (0.05); Shikoku (0.04) &Information (0.32); Retail (0.39) & 0.26 & 0.37 & 0.36\\ \hline
\end{tabular}
}
\begin{flushleft} 
The overexpression in terms of regions and sector-divisions in the five largest communities at the 2nd level, continued.
``Rank" refers to the rank among all the 2nd level subcommunities.
The percentage of nodes having a particular attribute is indicated in parentheses.
\end{flushleft}
\label{table_2nd_modular_B}
\end{table}


In addition, we present the overexpression of 99 sectors and 47 prefectures in the following subcommunities. 

\subsection{Five largest subcommunities}
Rank: 1

Overexpression of prefectures: Hokkaido; Tokyo; Fukui; Yamanashi; Tokushima; Kagawa; and Ehime

Overexpression of sectors: Information services; Financial products transaction dealers and futures
commodity dealers; Insurance institutions, including insurance agents brokers and services; Real estate
agencies; Real estate lessors and managers; Professional services, not elsewhere classified; and Miscellaneous living-related 
and personal services

Rank: 2

Overexpression of prefectures: Ibaraki; Tochigi; Gunma; Saitama; Chiba; Kanagawa; Nagano; Gifu; Shizuoka;
Aichi; Mie; Osaka; Hyogo; Nara; and Wakayama

Overexpression of sectors: Miscellaneous retail trade and Medical and other health services

Rank: 3

Overexpression of prefectures: Iwate; Tochigi; Kanagawa; Kyoto; Hiroshima; Fukuoka; Miyazaki; and Kagoshima

Overexpression of sectors: Medical and other health services

Rank: 4

Overexpression of prefectures: Hokkaido; Iwate; Tochigi; Gunma; Gifu; Aichi; Kyoto; Fukuoka; and Kagoshima

Overexpression of sectors: Manufacture of plastic products, except otherwise classified; Manufacture of business oriented machinery; Wholesale trade (machinery and equipment); Miscellaneous wholesale trade;
Real estate lessors and managers; and Services for amusement and recreation

Rank: 5

Overexpression of prefectures: Hokkaido; Chiba; Kanagawa; Gifu; and Aichi

Overexpression of sectors: Wholesale trade (general merchandise); Wholesale trade (machinery and equipment); and
Retail trade (machinery and equipment)

\subsection{Five largest subcommunities of the largest community:} 

Rank: 6

Overexpression of prefectures: Hokkaido; Aomori; Iwate; Akita; Niigata; Toyama; Ishikawa; Fukui; Okayama; Yamaguchi; Tokushima; Kagawa; Ehime; Kochi; Kumamoto; Oita; and Kagoshima

Overexpression of sectors: Retail trade (machinery and equipment)

Rank: 16

Overexpression of prefectures: Chiba; Osaka; and Wakayama

Over-expression of sectors: Construction work by specialist contractors, except equipment installation work;
Manufacture of textile products; Manufacture of iron and steel;
Road freight transport; Warehousing; Services incidental to transport; Wholesale trade (textile
and apparel); Wholesale trade (building materials, minerals and
metals, etc); and Waste disposal business

Rank: 25

Overexpression of prefectures: Hokkaido; Aomori; Akita; Toyama; Yamaguchi; Tokushima; Kochi; Saga; and Okinawa

Overexpression of sectors: Retail trade (machinery and equipment)

Rank: 38

Overexpression of prefectures:  Hokkaido; Iwate; Akita; Niigata; Toyama; and Fukui

Overexpression of sectors:  Retail trade (machinery and equipment)

Rank: 39

Overexpression of prefectures: Hokkaido; Shimane; Miyazaki; Kagoshima; and Okinawa

Overexpression of sectors: Manufacture of furniture and fixtures; Wholesale trade (machinery and equipment); Miscellaneous wholesale trade; and Eating and drinking places

\subsection{Five largest subcommunities of the 2nd largest community:}

Rank: 8

Overexpression of prefectures: Tokyo; Yamanashi; Nagano; and Okinawa

Overexpression of sectors: Road passenger transport; Water transport; Accommodations; Miscellaneous
living-related and personal services; Miscellaneous education and learning support; Employment and worker dispatching services; and Political, business and cultural organizations

Rank: 23

Overexpression of prefectures: Tokyo; Kanagawa; Kyoto; and Osaka

Overexpression of sectors: Manufacture of textile products; Manufacture of furniture and fixtures; Manufacture of leather tanning, leather products and fur skins; Miscellaneous manufacturing
industries; Wholesale trade (general merchandise); Wholesale trade (textile and apparel); Miscellaneous
wholesale trade; Retail trade (general merchandise); and Retail trade (woven fabrics, apparel, apparel accessories and notions)

Rank: 27

Overexpression of prefectures: Saitama; Aichi; and Osaka

Overexpression of sectors: Manufacture of food; Manufacture of production machinery; and Wholesale trade 
(food and beverages)

Rank: 28

Overexpression of prefectures: Saitama; Tokyo; Kanagawa; Kyoto; Osaka; and Hyogo

Overexpression of sectors: Manufacture of beverages, tobacco and feed; Wholesale trade 
(food and beverages); Retail trade (general merchandise); and Retail trade (food and beverage)

Rank: 29

Overexpression of prefectures: Ibaraki; Tochigi; Gunma; and Chiba

Overexpression of sectors: Agriculture; Forestry; Construction work, general including public
and private construction work; Manufacture of chemical and allied products; Miscellaneous wholesale trade; and
Miscellaneous retail trade

\subsection{Five largest subcommunities of the 3rd largest community:} 
Rank: 12

Overexpression of prefectures: Hokkaido; Miyagi; Tokyo; Aichi; Hiroshima; Kagawa; and Fukuoka

Overexpression of sectors: Construction work by specialist contractors, except equipment installation
work; Manufacture of fabricated metal products; Goods rental and leasing; Technical
services, not elsewhere classified; Employment and worker dispatching services; and Miscellaneous business services

Rank: 41

Overexpression of prefectures: Tokyo and Osaka

Overexpression of sectors: Real estate agencies; Real estate lessors and managers; Professional services, not elsewhere classified; and Advertising

Rank: 46

Overexpression of prefectures: Tokyo

Overexpression of sectors: Professional services, not elsewhere classified; Technical services, not elsewhere classified; Social insurance, social welfare and care services; Political, business and cultural organizations; and Local government services

Rank: 53

Overexpression of prefectures: Iwate; Akita; Tochigi; Niigata; Ishikawa; Nagano; Shizuoka; Aichi; Mie;
Nara; Wakayama; Okayama; Hiroshima; Yamaguchi; Kagawa; Fukuoka; Nagasaki; and Oita

Overexpression of sectors: Construction work, general including public and private construction work; and
Construction work by specialist contractors, except equipment installation work

Rank: 87

Overexpression of prefectures: Miyagi; Tochigi; Niigata; Toyama; Gifu; Shizuoka; Aichi; Mie; Tottori;
Shimane; Hiroshima; Tokushima; Kagawa;  Fukuoka; Saga; Oita; and Okinawa

Overexpression of sectors: Construction work, general including public and private construction work; and 
Construction work by specialist contractor, except equipment installation work

\subsection{Five largest subcommunities of the 4th largest community:} 
Rank: 1

Overexpression of prefectures: Hokkaido; Tokyo; Fukui; Yamanashi; Tokushima; Kagawa; and Ehime

Overexpression of sectors: Information services; Financial products transaction dealers and futures
commodity dealers; Insurance institutions, including insurance agents brokers and services; Real estate
agencies; Real estate lessors and managers; Professional services, not elsewhere classified; and Miscellaneous living-related 
and personal services

Rank: 5

Overexpression of prefectures: Hokkaido; Chiba; Kanagawa; Gifu; and Aichi

Overexpression of sectors: Wholesale trade, general merchandise; Wholesale trade (machinery and equipment); and
Retail trade (machinery and equipment)

Rank: 11

Overexpression of prefectures: Saitama; Kanagawa;and Ehime

Overexpression of sectors: Retail trade (machinery and equipment)

Rank: 17

Overexpression of prefectures: Hokkaido and Osaka

Overexpression of sectors: Manufacture of transportation equipment; Wholesale trade (machinery and
equipment); Retail trade (machinery and equipment); and Machine, etc. repair services, except otherwise
classified

Rank: 24

Overexpression of prefectures: Osaka

Overexpression of sectors: Retail trade (woven fabrics, apparel, apparel accessories and notions) and
Automobile maintenance services

\subsection{Five largest subcommunities of the 5th largest community:} 

Rank: 4

Overexpression of prefectures: Hokkaido; Iwate; Tochigi; Gunma; Gifu; Aichi; Kyoto; Fukuoka; and Kagoshima

Overexpression of sectors: Manufacture of plastic products, except otherwise classified; Manufacture of business oriented machinery; Wholesale trade (machinery and equipment); Miscellaneous wholesale trade;
Real estate lessors and managers; and Services for amusement and recreation

Rank: 22

Overexpression of prefectures: Hokkaido; Tochigi; Gunma; Chiba; Yamanashi; and Tokushima

Overexpression of sectors: Road freight transport; Miscellaneous retail trade; and Advertising

Rank: 26

Overexpression of prefectures: Nagano; Shizuoka; Aichi; and Mie

Overexpression of sectors: Manufacture of lumber and wood products, except furniture; Miscellaneous
manufacturing industries; Miscellaneous wholesale trade; Miscellaneous retail trade; Miscellaneous
education and learning support; and Machine, etc. repair services, except otherwise classified

Rank: 42

Overexpression of prefectures: Gifu and Osaka

Overexpression of sectors: Manufacture of business oriented machinery; Miscellaneous manufacturing
industries; Miscellaneous wholesale trade; Miscellaneous retail trade; Goods rental and leasing; and
Services for amusement and recreation

Rank: 50

Overexpression of prefectures: Kagoshima

Overexpression of sectors: Communications; Services incidental to the Internet;and Retail trade 
(machinery and equipment)

\clearpage
\section{Intra-link density of weighted links}
\label{sec:dense}

Fig~13 of the main text in the Comparison of industrial sectors Section shows the matrixes that represent the number of inter oand intra links between groups.
If we weight, i.e., using sales volume, the links and create the same matrixes for the weighted links, then
the matrixes can further indicate the agglomerative behaviour of groups.

\begin{figure}[h]
\includegraphics[width=0.8\textwidth]{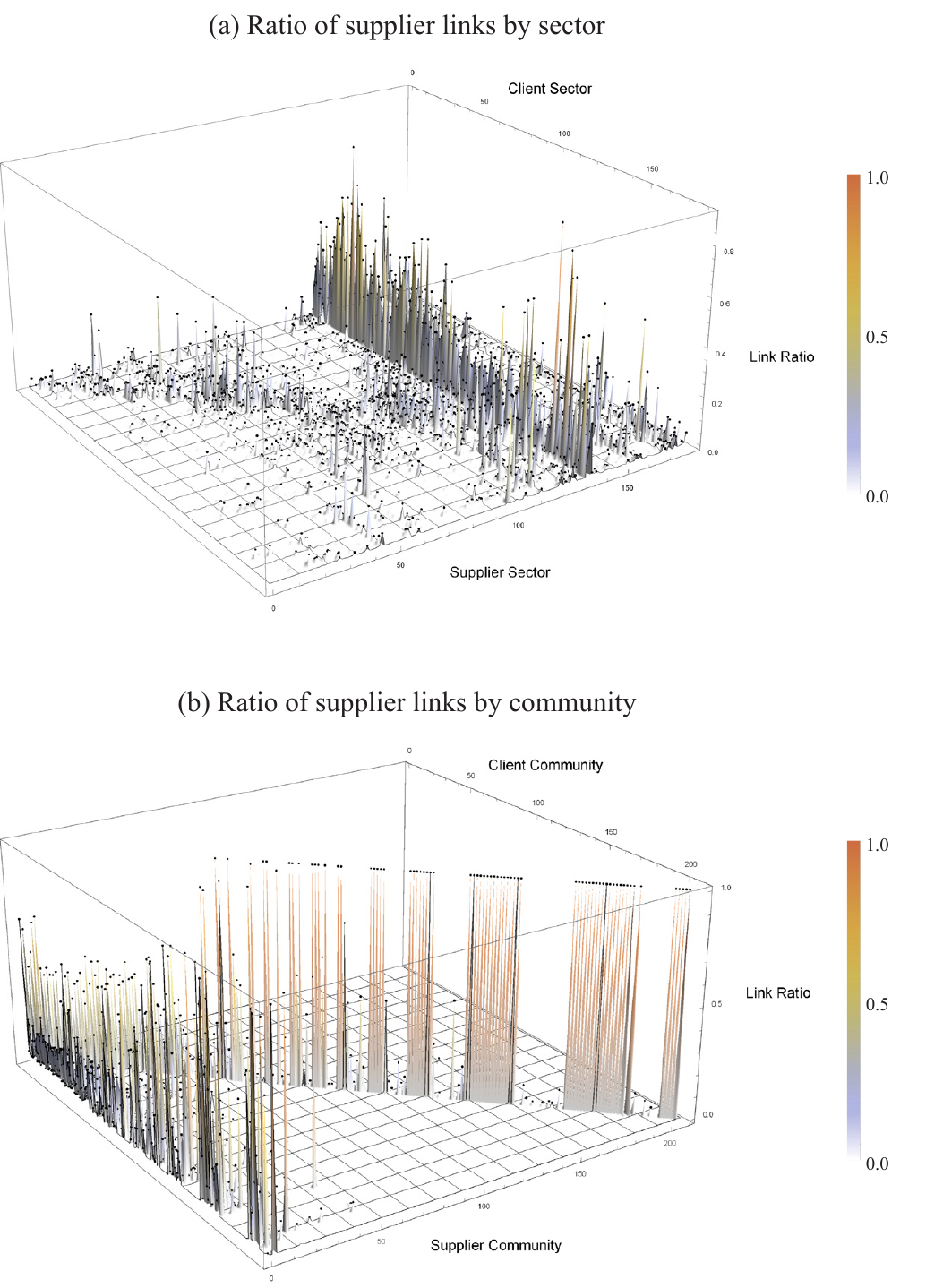}
\label{fig:ksec}
 \caption{{\bf Density of weighted links for the inter and intra-sectors of communities.}
These figures show the sales volume of the intergroups.
The top figure (a) shows the 3D plots of the industrial sectors.
The bottom figure (b) shows the 3D plots of the communities.}
 \label{fig:weightdense}

\end{figure}

Although the TSR data contains data on supplier and client relationships,
the sales volume for each relationship is not provided.
Therefore, we artificially add sales volume using the method proposed in \cite{Inoue2017}.
Each supplier's sales are proportionally divided into its clients' sales.
Here, we ignore supplier's sales to final consumers,
and the client's purchase from the supplier can be relatively estimated by using the sales of the clients as proxies.

Fig~\ref{fig:weightdense} provides the results.
The visualization of the industrial sectors has denser connections for wholesale and retailing than that for the number of links
shown in Fig~13 of the main text.
In addition, 
the visualization of the communities has denser connections to communities in the left rows than that shown in Fig~13 of the main text.
The overall ratio of internal sales volume,
i.e., (the total volume of the intra-group links)/(the total volume of all links) is
14.3\% for the industrial sectors and 55.9\% for the communities.

As a result, we obtain weaker connections between the intra-groups for both industrial sectors and communities.
The weaker intra-connections of these communities is understandable
because we do not use weighted links to detect the communities.
However, we do not expound on weighted community detection
to simplify the discussion in this paper.


\end{document}